\documentclass[%
notitlepage,
nofootinbib,
amsmath,
amssymb,
aps,
twocolumn
]{revtex4-1}

\usepackage{graphicx}
\usepackage{dcolumn}
\usepackage{bm}
\usepackage{mathtools}
\usepackage{amsmath}
\usepackage{amssymb}
\usepackage[hidelinks]{hyperref}
\usepackage{setspace}

\begin{document}

\title{A numerical exploration of first-order relativistic hydrodynamics}

\author{Alex Pandya}
\email{apandya@princeton.edu}
\author{Frans Pretorius}
\email{fpretori@princeton.edu}
 \affiliation{Department of Physics, Princeton University, Princeton, New Jersey 08544, USA.}

\date{\today}

\begin{abstract}
We present the first numerical solutions of the causal, stable relativistic Navier-Stokes equations as formulated by Bemfica, Disconzi, Noronha, and Kovtun (BDNK).
For this initial investigation we restrict to plane-symmetric configurations of a conformal fluid in Minkowski spacetime.  We consider evolution of three classes of initial data: a smooth (initially) stationary concentration of energy, a standard shock tube setup, and a smooth shockwave setup.
We compare these solutions to those obtained with a code based on the M\"uller-Israel-Stewart (MIS) formalism, variants of which are the common tools used today to model relativistic, viscous fluids. We find that for the two smooth initial data cases, 
simple finite difference methods are adequate to obtain stable, convergent solutions to the BDNK equations. 
For low viscosity, the MIS and BDNK evolutions show good agreement. At high viscosity the solutions begin to differ in regions with large gradients, and there the BDNK solutions can (as expected) exhibit violation of the weak energy condition. This behavior is transient, and the solutions evolve toward a hydrodynamic regime in a way reminiscent of an approach to a universal attractor.
For the shockwave problem, we give evidence that if a hydrodynamic frame is chosen so that the maximum characteristic speed of the BDNK system is the speed of light (or larger), arbitrarily strong shockwaves are smoothly resolved. Regarding the shock tube problem, it is unclear whether discontinuous initial data is mathematically well-posed for the BDNK system, even in a weak sense. Nevertheless we attempt numerical solution, and then need to treat the perfect fluid terms 
using high-resolution shock-capturing (HRSC) methods. When such methods can successfully evolve the solution beyond the initial
time, 
subsequent evolution agrees with corresponding MIS solutions, as well as the perfect fluid solution in the limit of zero viscosity. 
\end{abstract}

\maketitle

\section{Introduction} \label{sec:introduction}
In a modern interpretation, hydrodynamics can be thought of as 
a coarse-grained model of an underlying microscopic theory, allowing
for tractable study of certain macroscopic phenomena.
In that sense then hydrodynamics is not a single theory, but a hierarchy of theories
that successively include more details and properties
of the underlying microphysics (see e.g.~\cite{Kovtun_2012,Romatschke_2019}).
The leading order model (zeroth order in a gradient expansion) is applicable to matter in 
local thermodynamic equilibrium,
characterized by basic material properties such as energy density and temperature,
and subject to evolution equations consistent with stress-energy conservation
(the Euler equations), conservation of particle number
for baryons, etc. At next-to-leading (first) order, effects associated
with deviations from equilibrium appear, such as viscous
dissipation due to velocity gradients, or heat conduction
due to thermal gradients. The corresponding statement of stress-energy conservation
is captured by the Navier-Stokes equations.

Despite the simple physical principles that underlie these 
hydrodynamic theories, the equations are non-linear, and
even exhibit complicated phenomena such as turbulence, and singular
behavior (discontinuities) in some shockwaves. Singularities are often a problem
for the predictability of a theory, though 
for the Euler equations, requiring stress-energy conservation and consistency
with the second law of thermodynamics
is adequate to allow for unique weak-form solutions 
that accurately capture the behavior outside of the discontinuity \cite{1967pswh.book.....Z}.
In other words, the details of the microphysics that would ostensibly
resolve the discontinuity seems irrelevant on large scales, and
remarkably, the Euler equations reflect this, despite
a complete breakdown of the small-gradient assumption 
that would otherwise justify them as a sound
mathematical model of the corresponding physical phenomenon.

Historically, the success of hydrodynamics as a 
model of the dynamics of macroscopic distributions of matter seemed to fail at first order for 
relativistic theories, as originally formulated by Eckart \cite{Eckart_1940} in 1940,
and a different variant by Landau and Lifshitz in the 1950's \cite{Landau_1987}.
A problem recognized early on is that the resultant relativistic
Navier-Stokes equations are parabolic, inconsistent with
causality as defined by the postulates of relativity. A reasonable assumption
would have been that this simply implies a limited range of scenarios where the relativistic
Navier-Stokes equations should be expected to provide accurate predictions. However, 
that notion was dramatically disproven by
Hiscock and Lindblom in 1985~\cite{Hiscock_1985}, when they showed these theories
do not admit stable equilibrium solutions for reasonable forms of matter,
even in non-relativistic settings.

To address the issues of hyperbolicity and causality, in the 1960's
M\"uller \cite{Muller_1967}, and subsequently Israel and Stewart~\cite{Israel:1976tn,Israel_1979},
showed that the inclusion of second-order terms may be able to 
yield a more suitable theory.  Though the additional terms 
significantly complicate nonlinear analysis, the theory was later shown 
to be stable, causal, and hyperbolic when linearized about equilibrium 
\cite{Hiscock_1983}, motivating its use over the theories of Eckart and 
Landau-Lifshitz. As a result, the so-called M\"uller-Israel-Stewart (MIS) 
theories 
are behind essentially all current numerical efforts to model relativistic dissipative fluids (see~\cite{Romatschke_2019} for a comprehensive review). 
Applications of contemporary interest 
include modeling relativistic heavy ion collisions~\cite{Romatschke_2019}, 
neutron star dynamics~\cite{Stergioulas:2003yp,Duez:2018jaf}, 
early universe cosmology~\cite{Brevik:2014cxa}, plasma physics~\cite{1989rfmw.book.....A}, black hole accretion~\cite{Done:2007nc,2010tbha.book.....A} and relativistic jets~\cite{Blandford:2018iot}.

Though highly successful in the above mentioned applications, there are a few 
issues with MIS theories. One is that of aesthetics: it seems rather
excessive to need all the complexity of the second-order theory
(with over a dozen new transport coefficients that appear relative
to first order~\cite{Romatschke_2019}) if only the first-order terms are expected to be
relevant in a given problem. As such, many practitioners use a
truncated version of MIS (as we do here for the comparison model,
discussed more in Sec.~\ref{sec:MIS_covariant} below); though somewhat ad-hoc, this
is justifiable in scenarios where second-order effects are small. 
Another problem with MIS-type theories is they generically do
not admit solutions describing high Mach number
strong shocks, even in a weak sense~\cite{Olson_1990_shocks,Geroch_1991}.
Though one might argue this is not surprising for a theory based on 
a gradient expansion, and it is likely just the simplicity of the Euler
equations that the latter can be ``fixed'' in this regard, it would still mean
that beyond zeroth order, relativistic hydrodynamics breaks down as a
predictive theory when strong shocks form.

There have been other proposals
to resolve the problems with traditional first-order relativistic hydrodynamics (in particular \cite{1955PhT.....8R..24L}),
though just within the past decade has 
a revolution in understanding the source of the latter's pathology arisen,
giving a clear and systematic approach to constructing well-posed
first-order theories. Building on earlier work by V\'an and Bir\'o~\cite{Van:2011yn} and
Freist\"uhler and Temple~\cite{2014RSPSA.47040055F,2017RSPSA.47360729F,2018JMP....59f3101F},
the key insight by Bemfica, Disconzi and Noronha~\cite{Bemfica_2018},
and expanded upon by Kovtun~\cite{Kovtun_2019}, was recognizing how
the choice of the {\em hydrodynamic frame} influences the
hyperbolicity of the underlying equations (in this paper we will
often simply use ``frame'' when referring to the hydrodynamic frame,
and when we use ``reference frame'' or ``rest frame'' we mean
a coordinate (Lorentz) frame). 

Eckart and Landau-Lifshitz already knew
that the hydrodynamic variables do not have unique definitions outside of equilibrium.
The hydrodynamic frame is then essentially the choice of definition
of a complete set of fundamental variables, such as the flow four-velocity $u^a$, energy density $\epsilon$, and particle
number density $n$ (or equivalent replacements of thermodynamic quantities using the equation
of state), and how they relate to the stress-energy tensor $T^{ab}$, particle four-current $J^a$ and various transport 
coefficients through a series of constitutive relations.
For example, out of equilibrium, the particle number flux 4-velocity vector $u_N^a$ does not need
to be tangent to the energy flux 4-velocity $u_E^a$ (defined as an eigenvector of $T^{ab}$); amongst other choices, 
Eckart used a frame where $u^a=u_N^a$, while Landau and Lifshitz used one where $u^a=u_E^a$.
The choice of frame constrains the set of transport coefficients appearing in $T^{ab}$ and $J^{a}$, and by considering field
redefinitions one can determine how some vary and others are invariant under changes of hydrodynamic frame~\cite{Kovtun_2019}. 
More importantly for our discussion, the choice of frame also affects the character
of the partial differential equations (PDEs) in the resultant Navier-Stokes and various
charge conservation equations: a judicious choice of frame allows
for a well-posed, strongly hyperbolic system of PDEs with causal propagation 
speeds~\cite{Bemfica_2018,Bemfica:2019hok,Bemfica:2019knx,Bemfica:2020zjp,Hoult_2020}.

The purpose of this paper is to report on initial results implementing
the relativistic viscous hydrodynamics theories of Bemfica, Disconzi, Noronha and Kovtun (BDNK) 
in a numerical solution scheme, which to our knowledge has not been done before.
Given the decades of research into developing stable codes to solve the
relativistic Euler equations, and MIS-inspired schemes to model dissipative
corrections, it may seem like it would be a trivial 
process to re-tool one of these codes to solve the BDNK systems. Indeed, 
one of the results from our work is that standard methods {\em can}
straight-forwardly be adapted to the BDNK equations, at least for the scenario
studied here: a conformal fluid restricted to planar symmetry in Minkowski spacetime. 
However, that is not {\em a priori} an obvious 
conclusion for a few reasons. The main one is related to what portion of the stress tensor $T^{ab}$ contains
the principle parts of PDEs that govern the equations' character (here hyperbolic), and what that
implies for numerical solution.
The Euler equations are
most commonly written in flux-conservative form, allowing for the application of
Godunov-type methods to deal with discontinuities that form in many scenarios
of interest. Such techniques essentially assume a discontinuity is present
at each cell interface, and solve an exact or approximate Riemann problem
at each interface to update cell averages of the fluid variables at each time step.
As illustrated in more detail in Sec. \ref{sec:MIS_covariant}, the addition of dissipative terms via the 
MIS approach does not alter the basic
structure of the hydrodynamic evolution, as the higher order corrections to the 
stress tensor are elevated to the status of new fundamental variables with their own evolution equations, and only couple
to the Euler equations as lower order source terms. 

For the BDNK equations this
is not the case: the principle parts of the PDEs are now entirely determined
by the viscous part of the stress tensor, with the Euler terms relegated to lower
order. 
Stable, convergent numerical solution schemes mirror the proofs of the well-posedness
of continuum equations: they must be tailored to the structure
of the principle parts of the PDEs, and then (for the most part) the lower order
terms will not adversely affect the numerical evolution. The difficulty contemplating
taking this route for a BDNK system is that for 
equilibrium states, which will generically
be present in at least parts of the domain, the viscous terms
are identically zero, and near equilibrium evolution is governed by the lower
order Euler terms. Since the latter by themselves are also hyperbolic,
this should not be a problem for smooth flows. 
Though when shocks form, even if in principle viscosity is able to smooth them,
the scale over which the fluid profile smoothly transitions from one state to another
may be too small to resolve in practice. Also, nothing prevents
one from putting in non-smooth initial data, and certain applications
effectively require this (e.g. the moment of collision in a binary neutron
star merger). Thus it seems important to maintain the ability
of a numerical scheme to stably evolve non-smooth data when dissipative 
effects are included, but it is unclear whether the analogue
of the Riemann problem makes mathematical sense for a theory governed by second-order PDEs such as BDNK\footnote{For the non-relativistic Navier-Stokes equations
similar reasoning holds. Another difficulty in that case is the dissipative
terms make the equations parabolic, which can impose severe
time-stepping restrictions for stable evolution; methods
have been developed to alleviate this, such as those shown in~\cite{Kurganov_2000}.}.

In the remainder of this introduction, we outline the rest of the paper, and give a brief 
summary of our main results.

After a more general discussion of the gradient expansion in Sec. \ref{sec:grad_exp},
in Secs.~\ref{sec:Euler}, \ref{sec:RNS_covariant} and \ref{sec:MIS_covariant} we describe 
the perfect fluid, BDNK and truncated
MIS systems we consider here, respectively. Beyond demonstrating stable evolution of
the BDNK equations, one goal is to compare evolution of identical
initial data using these three different theories for a select set of problems, and
identifying in what regimes they agree.
For the dissipative schemes we also investigate some self-consistent diagnostic
measures (described in Sec.~\ref{sec:diagnostics})
to check whether the state has evolved to a regime where the results 
should not be trusted, even if there is no breakdown or other
apparent issue with the numerical solution. For simplicity
in this first study we restrict to a conformal fluid, and
planar symmetry in Minkowski spacetime (i.e. (1+1)D evolution), and in 
Sec.~\ref{sec:slab} give the explicit form of the three sets of equations we will
solve numerically. In Sec.~\ref{sec:numerical_method} we describe the
numerical methods we employ. For concreteness, we will choose
parameters of the test problems to mimic conditions relevant
to heavy ion collisions; we discuss this and the units we use
in Sec.~\ref{sec:phys}.

We present results in Sec.~\ref{sec:results}, one for an
initially static fluid with a Gaussian distribution
for the energy density, the second
a standard shock tube problem with discontinuous
initial data, and the third smooth initial data transitioning
between an upstream supersonic flow and a downstream
subsonic flow.

For the Gaussian initial data, the perfect fluid eventually
develops shocks, while for BDNK and MIS with non-zero viscosity
the fluid variables remain smooth for the length of the simulations.
The dissipative schemes show similar results for low viscosity,
but begin to differ at high viscosity. In that regime
the BDNK solutions develop regions where the weak energy condition
is violated, as expected when the gradient terms in the
stress energy tensor become large~\cite{Bemfica_2018}. 
Interestingly, though the resultant solutions 
then are markedly different from the corresponding MIS solutions,
or between two BDNK solutions obtained with different hydrodynamic frames
(all having started with identical initial data),
these  ``non-hydrodynamic'' features decay away exponentially,
and the solutions soon closely resemble each other again.
This is reminiscent of so-called universal attractor behavior found
to be present in beyond-ideal theories modeling Bjorken flow 
~\cite{Heller:2015dha,Romatschke:2017vte} (a flow that seems to describe
the leading order phase of expansion of a
quark-gluon plasma formed in an ultra-relativistic heavy ion collision).

For the shock tube test, similar energy condition violations occur near
the initial time, but soon afterward, both for MIS and BDNK,
the evolution approaches a state that looks like a smoothed
version of the perfect fluid case, with the diagnostics
suggesting the dissipative corrections have become small from
the perspective of the gradient expansion.

Regarding shockwaves in viscous hydrodynamics, 
as mentioned above, there are theorems that in MIS-type relativistic theories solutions
do not exist for sufficiently strong (high Mach number) shocks, even in a weak sense. This
is disconcerting, and considered by some a significant shortcoming
of such theories~\cite{Olson_1990_shocks,Geroch_1991,2014RSPSA.47040055F}. 
However, the nature of the proofs are more
suggestive of a failure of hyperbolicity than some intrinsic inadequacy
of relativistic dissipative hydrodynamics: the limiting upstream
velocity above which shock solutions cease to exist is precisely
when the largest upstream characteristic speed of the system
becomes zero in the observer's reference frame\footnote{The characteristic speeds of the PDEs governing beyond-ideal hydrodynamic
theories generally do not coincide with the sound speed of the fluid.}.
For such supersonic
flows, information about the downstream state cannot
be propagated upstream, arguing for the presence of discontinuities in
the flow at the shock front. This is what happens with the Euler equations,
and sensible weak-form solutions can be derived there, giving
the Rankine-Hugoniot jump conditions. For the Newtonian Navier-Stokes equations,
discontinuities are not inevitable, as the equations are parabolic and information
can always be propagated upstream regardless of the flow speed;
intuition suggests the dissipative terms should smooth
out the shock front, and this is confirmed by numerical solution, even
though in some cases shock properties do not
match experiments very well (see e.g.~\cite{1976JFM....74..497A}). It would be
a curious circumstance if viscosity failed to ``work'' in this sense in relativity,
but would be acceptable if weak form solutions still existed; that they
do not for MIS-type theories is a much more severe problem, for, as mentioned,
it implies failure of the Cauchy problem and subsequent loss of predictability. 

As we demonstrate here, for the shockwave test case similar problems 
can be present within BDNK theories. However, we also show
that this is tied to the choice of hydrodynamic frame, and we can
choose one where the trends indicate arbitrarily strong shocks
can be smoothly resolved. This is consistent with the
above theorems~\cite{Olson_1990_shocks,Geroch_1991} in that 
the ``good'' frames for resolving shocks are those where
the maximum characteristic speed is the speed of light (or larger).
This is also consistent with the work
of Freist\"uhler and Temple~\cite{2014RSPSA.47040055F}, who demanded existence
of arbitrarily strong shock solutions as a defining criterion
for the class of first-order relativistic theories they proposed.
In doing so, they had to abandon the restriction that entropy
production is positive along all gradients, but found that
violations of the second law actually do not occur along shock profiles. 
We similarly observe positive entropy production along shocks,
though given that we are using a conformal fluid (there is a simple
one-to-one relationship between entropy density $s$ and energy density
$\epsilon$), and the shock
smoothly transitions between flows which asymptotically approach the corresponding perfect fluid states (by conservation
of stress-energy), this is somewhat of a trivial conclusion in our case.
That is not to suggest that sufficiently
far out of equilibrium regimes do not exist where
BDNK theories could show violation of the second law. However,
one could view such pathological evolution as a ``feature''
of these theories, providing an additional diagnostic---similar to the weak energy condition violation---to tell when the fluid is outside the realm where only first-order
dissipative corrections are adequate to describe its dynamics.

As we completed this work, a paper by Freist\"uhler~\cite{freistuhler2021nonexistence}
appeared on the archive that
proves some results related to strong shocks within the BDNK system,
and likewise seems consistent with the above discussion. 
In the results below we will mention where the particular examples we present fall within
the characterization of the BDNK frames introduced in~\cite{freistuhler2021nonexistence}.

We conclude in Sec.~\ref{sec:conclusion} with a discussion
of potential follow up work. We leave the explicit form of the primitive
variable recovery and numerical algorithm for our BDNK scheme to Appendices \ref{sec:RNS_prim_var_recovery}
and \ref{sec:RNS_pseudocode} respectively, a listing of the steady state equations relevant to the shockwave
problem to Appendix \ref{app_ss},
and some convergence tests to Appendix~\ref{sec:convergence_tests}.

\section{The gradient expansion}\label{sec:grad_exp}
In this section we begin by reviewing hydrodynamics from the perspective of a gradient
expansion. Then in Sec.~\ref{sec_conf} we describe simplifications that result for a conformal
fluid, followed by details specific
to the zeroth, first and second-order theories we consider here in Secs.~\ref{sec:Euler},~\ref{sec:RNS_covariant} and~\ref{sec:MIS_covariant} respectively.

Relativistic fluid models are typically defined in terms of two conserved currents: the stress-energy tensor $T^{ab}$, which arises as a result of spacetime translation invariance, and a U(1) conserved current $J^{a}$, coming from the local conservation of the number of particles (baryons) \cite{Kovtun_2012,Bemfica_2020}. These currents
are functions of a set of \textit{hydrodynamic variables}: the energy density $\epsilon$, the baryon number density $n$, isotropic rest-frame pressure $P$, the flow four-velocity $u^{a}$, chemical potential $\mu$, 
temperature $T$, etc., that define the macroscopic state of the fluid.
The corresponding conservation laws are
\begin{align}
\nabla_{a} T^{ab} &= 0  \label{eq:T_ab_conservation_law} \\
\nabla_{a} J^{a}  &= 0, \label{eq:J_conservation_law}
\end{align}
where $\nabla_{a}$ is the covariant derivative compatible with the spacetime metric $g_{ab}$, which we take to have the ``mostly-plus'' signature $(-+++)$. This 
gives $d+1$ equations constraining the dynamics of $T^{ab}$ and $J^{a}$, where $d$ is the dimension of the 
spacetime (we consider $d=4$ here). A defining feature of hydrodynamics is that these $d+1$ equations
are assumed to be sufficient to predict the unique evolution of the state of the fluid from given initial data.
This is only possible because not all the hydrodynamic variables are independent; the additional constraints needed to close the system come from the thermodynamic \textit{equation of state}, which characterizes microphysical properties of the particular fluid under consideration. That further implies there is freedom in choosing a subset of these as independent variables that will be explicitly solved for; in the description below we will mainly use $\epsilon$, $n$, 
and $u^{a}$ (this is a common choice for astrophysical applications).

For fluids in local thermodynamic equilibrium, or when dissipative effects are negligible, $T^{ab}$ and $J^{a}$
are simply algebraic functions of the hydrodynamic variables, and the interpretation of these variables is unambiguous. Outside of equilibrium, however, this is no longer true.
Nevertheless, it is conventional to assume that $T^{ab}, J^{a}$ may be still be parameterized by the hydrodynamic variables provided the fluid is sufficiently close to equilibrium, though one must now also consider combinations of the hydrodynamic variables and their derivatives.
In particular, near equilibrium one assumes one can express the conserved currents in terms of a \textit{gradient expansion}
\begin{equation} \label{eq:gradient_expansion}
\begin{aligned}
T^{ab} &= T^{ab}_{(0)} + T^{ab}_{(1)} + T^{ab}_{(2)} + ... \\
J^{a}  &= J^{a}_{(0)}  + J^{a}_{(1)}  + J^{a}_{(2)} + ...,
\end{aligned}
\end{equation}
where the zeroth-order terms $T^{ab}_{(0)}, J^{a}_{(0)}$ are the equilibrium case considered previously. At first order, $T^{ab}_{(1)}, J^{a}_{(1)}$ depend linearly on first derivatives of the variables, i.e. $\nabla^{a} u^{b}$, $\nabla^{a} \epsilon$ and
$\nabla^{a} n$.  At second order and above, one counts higher order gradients and products of lower order gradients on the same footing; for example, both $\nabla^{a} \nabla^{b} \epsilon$ and $(\nabla^{a} \epsilon) (\nabla^{b} \epsilon)$ would appear in the second-order contribution to the stress-energy tensor, $T^{ab}_{(2)}$.

It is impractical to work with $T^{ab}, J^{a}$ in (\ref{eq:gradient_expansion}) up to high order in gradients\footnote{And in fact, the series likely has a zero radius of convergence at infinite order~\cite{Heller:2013fn,Buchel:2016cbj,Denicol:2016bjh,Heller:2016rtz}.} , so we will always truncate them at some order $k$. In these truncated expressions, following the notation of~\cite{Hoult_2020}, we will use the subscript $k$ (without parentheses) to define a quantity computed up to and including $k$th order gradients, e.g.
\begin{equation}
T^{ab}_{k} = T^{ab}_{(0)} + T^{ab}_{(1)} + ... + T^{ab}_{(k)},
\end{equation}
where the terms with subscript $(k)$ (with parentheses) denote a term entirely at $k$th order.  It will be useful to also define the dissipative corrections to the stress-energy tensor and particle current:
\begin{align}
T^{ab}_{k} &= T^{ab}_{(0)} + \pi^{ab}_{k} \label{eq:pi_defn}\\
J^{ab}_{k} &= J^{ab}_{(0)} + \xi^{ab}_{k},
\end{align}
which include all gradient corrections to the equilibrium stress-energy tensor and particle current, e.g. $\pi^{ab}_{k} = T^{ab}_{(1)} + ... + T^{ab}_{(k)}$.

At this point it is possible to define $T^{ab}, J^{a}$ by writing them as linear combinations of all possible gradient terms of the hydrodynamic variables $\{\epsilon, n, u^{a}\}$ up to $k$th order.  As $k$ increases, however, the number of possible terms grow rapidly and the need for a bookkeeping system becomes apparent.  It is conventional to begin by decomposing $T^{ab}, J^{a}$ in terms of $u^{a}$, which is taken to be timelike, $u_{c} u^{c} = -1$.  
Then, without loss of generality \cite{Kovtun_2012}
\begin{align}
T^{a b} &= \mathcal{E} u^{a} u^{b} + \mathcal{P} \Delta^{a b} + (\mathcal{Q}^{a} u^{b} + \mathcal{Q}^{b} u^{a}) + \mathcal{T}^{a b} \label{eq:decomp_T_ab} \\
J^{a} &= \mathcal{N} u^{a} + \mathcal{J}^{a}, \label{eq:decomp_J_a}
\end{align}
where $\mathcal{E}, \mathcal{P}, \mathcal{N}$ are scalars; $\mathcal{Q}^{a}, \mathcal{J}^{a}$ are vectors transverse to $u^{a}$ (i.e. $u_{a} Q^{a} = u_{a} J^{a} = 0$); $\mathcal{T}^{ab}$ is a symmetric transverse traceless tensor ($u_a \mathcal{T}^{ab} = g_{ab} \mathcal{T}^{ab} = 0$); and the symmetric tensor
\begin{equation}
\Delta^{ab} \equiv g^{ab} + u^{a} u^{b}
\end{equation}
projects onto the space transverse to the fluid velocity ($u_a \Delta^{ab} = 0$). 
In terms of $T^{ab}, J^{a}$, these quantities are defined by
\begin{multline} \label{eq:tensor_decomp_definitions}
\mathcal{E} = u_c u_d T^{cd}, ~~~
\mathcal{P} = \frac{1}{d-1} \Delta_{cd} T^{cd}, ~~~
\mathcal{Q}_{a} = - \Delta_{a c} u_d T^{cd} \\
\mathcal{N} = -u_c J^{c}, ~~~
\mathcal{J}_{a} = \Delta_{a c} J^{c}, ~~~ 
\mathcal{T}^{ab} = T^{<ab>}.
\end{multline}
The angle brackets are shorthand for
\begin{multline} \label{eq:TT_part}
X^{<ab>} = \frac{1}{2} \Big( \Delta^{ac} \Delta^{bd} X_{cd} + \Delta^{ac} \Delta^{bd} X_{dc} \\
- \frac{2}{d-1} \Delta^{ab} \Delta^{cd} X_{cd} \Big),
\end{multline}
which gives the transverse traceless part of a general rank-two tensor $X^{ab}$ ($u_a X^{<ab>} = g_{ab} X^{<ab>} = 0$).

Specifying a fluid theory at order $k$ amounts to replacing (\ref{eq:tensor_decomp_definitions}) with a set of \textit{constitutive relations} defining $\{\mathcal{E}_{k}, \mathcal{P}_{k}, \mathcal{Q}^a_{k}, \mathcal{T}^{ab}_{k}, \mathcal{N}_{k}, \mathcal{J}^a_{k}\}$ in terms of the hydrodynamic variables $\{\epsilon, n, u^{a}\}$, the spacetime metric $g^{ab}$, and their gradients up to order $k$.

\subsection{Conformal fluids}\label{sec_conf}
Before writing down the constitutive relations at zeroth, first, and second order---corresponding to the relativistic Euler, BDNK, and MIS equations respectively---we will restrict our attention to a fluid with an underlying conformal symmetry ($g_{ab} T^{ab} = 0$) and no conserved baryon current ($J^{a} = 0$).  These assumptions yield a significant simplification to the stress-energy tensor at higher orders of the gradient expansion, and allow us to more easily make contact with established results from the relativistic heavy ion collision community, 
which often uses a viscous conformal fluid as a toy model for quark-gluon plasma\footnote{Though QGP is often far from conformal in heavy-ion collisions \cite{Shuryak_2009}, quantum chromodynamics (QCD) is nearly conformal at sufficiently high temperatures \cite{Baier_2008}.} (QGP).

A straightforward calculation shows that tracelessness of the perfect fluid $T^{ab}$ (see (\ref{eq:PF_T_ab}) below) requires the equation of state relating the fluid pressure $P$ to the energy density $\epsilon$ to be $P = \epsilon/3$ (for $d=4$).
This result also implies that $\epsilon = \epsilon_0 T^{4}$, where $T$ is the temperature and $\epsilon_0$ is a dimensionful constant whose value should be derived from the thermodynamics of the substance being modeled.  

A simple example of a conformal fluid is a gas of free, massless particles, such as a free photon gas, or a perfect fluid with the so-called ultrarelativistic equation of state \cite{Neilsen_2000} $P = (\Gamma - 1) \epsilon$ with $\Gamma = 4/3$.

\subsection{Zeroth-order hydrodynamics: relativistic Euler equations}\label{sec:Euler}
Since we are considering a conformal fluid with no conserved particle number $n$, $\{\epsilon, u^{a}\}$ are the only hydrodynamic variables that will appear in the constitutive relations.  Using the velocity decomposition for $T^{ab}$ (\ref{eq:decomp_T_ab}), one sees that the hydrodynamic variables alone cannot form a transverse vector or a transverse traceless tensor, so $\mathcal{Q}^a = \mathcal{T}^{ab} = 0$.  We are left with only the scalars $\mathcal{E}, \mathcal{P}$, each of which must be a function of $\epsilon$.  An observer comoving with a fluid will see a rest frame energy density $\epsilon$ and isotropic pressure $P$ $(=\epsilon/3)$, requiring $\mathcal{E}=\epsilon$ and $\mathcal{P}=P$ in (\ref{eq:decomp_T_ab}). Thus
\begin{equation}
T^{ab}_{(0)} = \epsilon u^{a} u^{b} + P \Delta^{ab} \label{eq:PF_T_ab},
\end{equation}
which is the stress-energy tensor for a perfect (ideal) fluid.  Combining (\ref{eq:PF_T_ab}) with (\ref{eq:T_ab_conservation_law}) yields the relativistic Euler equations, which govern the time evolution of an inviscid fluid in local thermodynamic equilibrium.

\subsection{First-order hydrodynamics: relativistic Navier-Stokes equations} \label{sec:RNS_covariant}

\subsubsection{First-order constitutive relations}

At first order in the gradient expansion, one must now incorporate derivatives of the hydrodynamic variables into the constitutive relations defining $\mathcal{E}, \mathcal{P}, \mathcal{Q}^a, \mathcal{T}^{ab}$, replacing (\ref{eq:tensor_decomp_definitions}).  For a conformal fluid without a conserved baryon number $n$, the only allowed first-order terms are the scalars $\nabla_{c} u^{c}, u^c \nabla_c \epsilon$,  the transverse vectors $\Delta^{a c} \nabla_{c} \epsilon, u^c \nabla_c u^a$, and the shear tensor $\sigma^{ab} \equiv \nabla^{<a} u^{b>}$.  One can then show that the following are the complete set of linear combinations of these terms that arise at first order~\cite{Romatschke_2019}
\begin{equation} \label{eq:first_order_constitutive_relations}
\begin{aligned}
\mathcal{E}_{1} &= \epsilon + \mathcal{A}_{1} \\
\mathcal{A}_{1} &\equiv \frac{3 \chi}{4 \epsilon} u^c \nabla_c \epsilon + \chi \nabla_c u^c \\
\mathcal{Q}^{a}_{1} &= \frac{3 \lambda c_s^2}{4 \epsilon} \Delta^{ac} \nabla_c \epsilon + \lambda u^c \nabla_c u^a \\
\mathcal{T}^{ab}_{1} &= -2 \eta \sigma^{ab},
\end{aligned}
\end{equation}
where $\mathcal{P}_{1} = \mathcal{E}_{1}/3$ comes from the requirement that $T^{ab}$ be trace free, and $c_s^2 \equiv d P/d \epsilon = 1/3$ is the square of the sound speed for a conformal fluid.  The coefficients $\chi, \lambda, \eta$ are gradient-free functions of the hydrodynamic variables, and will be discussed in detail in the next subsection.

Before moving on, it will be useful to define the dissipative correction tensor at first order, $\pi^{ab}_{1}$ (\ref{eq:pi_defn}) corresponding to (\ref{eq:first_order_constitutive_relations}):
\begin{equation} \label{eq:pi_first_order}
\pi^{ab}_{1} = T^{ab}_{(1)} = \mathcal{A}_{1} \Big[ u^a u^b + \frac{\Delta^{ab}}{3} \Big] + (\mathcal{Q}^{a}_{1} u^b + \mathcal{Q}^{b}_{1} u^{a}) + \mathcal{T}^{ab}_{1}.
\end{equation}
In summary, $T^{ab}$ up to first order is defined by inserting (\ref{eq:first_order_constitutive_relations}) into (\ref{eq:decomp_T_ab}), or equivalently by inserting (\ref{eq:PF_T_ab}) and (\ref{eq:first_order_constitutive_relations})-(\ref{eq:pi_first_order}) into (\ref{eq:pi_defn}).

\subsubsection{First-order transport coefficients} \label{sec:first_order_coeffs}

The coefficients $\chi, \lambda, \eta$ are often referred to as \textit{transport coefficients}, and their particular functional forms depend both on the
choice of hydrodynamic frame, and physical properties of the underlying microscopic theory to which the fluid model is a long-wavelength approximation.
The coefficients $\chi, \lambda$ are not usually named, but in this case control the size of gradient corrections to the energy density ($\mathcal{A}_{1} \propto \chi$) and heat flow ($\mathcal{Q}^{a}_{1} \propto \lambda$) respectively\footnote{In \cite{Bemfica_2020} the coefficients $\lambda, \chi$ are replaced with relaxation times $\tau_{Q}, \tau_{\epsilon}, \tau_{P}$.  The requirement that $T^{a}_{~a} = 0$ for a conformal fluid forces $\tau_{P} = \frac{\tau_{\epsilon}}{3}$, and comparison of the tensor in \cite{Bemfica_2018} with that of \cite{Bemfica_2020} implies $\tau_{\epsilon} = \frac{3 \chi}{4 \epsilon}$ and $\tau_{Q} = \frac{3 \lambda}{4 \epsilon}$.}.
The remaining coefficient, $\eta$, is the \textit{shear viscosity} and determines the extent to which the fluid responds to trace-free gradients in the flow velocity $u^{a}$ ($\mathcal{T}^{ab}_{1} \propto \eta \nabla^{<a} u^{b>}$).  The fluid's response to the trace of the velocity gradient ($\nabla_{c} u^{c}$) determines its reaction to expansion or contraction, and can appear in various parts of the dissipative correction tensor; its contribution to the isotropic (trace) part can be thought of as a contribution to the fluid pressure, and is called the \textit{bulk viscosity} with coefficient $\zeta$.  The fact that a conformal fluid's stress-energy tensor is trace free implies that $\zeta = 0$, which is why $\zeta$ does not appear 
in (\ref{eq:first_order_constitutive_relations}).

Here we adopt the following 3-parameter $(\eta_0, \lambda_0, \chi_0)$ family
of transport coefficients,
\begin{equation} \label{eq:first_order_transport_coeffs}
\eta    \equiv \eta_0 \epsilon^{3/4}, ~~~ \lambda \equiv \lambda_0 \epsilon^{3/4}, ~~~ \chi    \equiv \chi_0 \epsilon^{3/4},
\end{equation}
where $\eta_0$ is a free parameter that largely determines the amount 
of dissipation in the fluid, and $\lambda_0, \chi_0$ are constants 
controlling the hydrodynamic frame. In \cite{Bemfica_2018}, existence
and uniqueness of solutions, causality, and linear stability about
equilibrium were proven provided the transport coefficients obey the 
following constraints: $\eta_0 > 0$, $\chi_0 = a_1 \eta_0$, and
$\lambda_0 \geq \frac{3 \eta_0 a_1}{a_1 - 1}$, with $a_1 \geq 4$. 
Here we take $\eta_0 > 0$ and consider two choices of hydrodynamic frame
\begin{equation} \label{frames}
\begin{aligned}
{\tt A:} \ \ \ & (\lambda_0, \chi_0) = \bigg( \frac{25 \eta_0}{3}, \, \frac{25 \eta_0}{2} \bigg),\\
{\tt B:} \ \ \ & (\lambda_0, \chi_0) = \bigg( \frac{25 \eta_0}{7}, \, \frac{25 \eta_0}{4} \bigg),
\end{aligned}
\end{equation}
which can be shown to satisfy the above constraints\footnote{The Eckart and Landau-Lifshitz theories instead choose $\chi = 0$ and $\lambda = \chi = 0$, respectively \cite{Kovtun_2012}; as mentioned in the introduction, these choices lead to acausal equations of motion with unstable equilibrium states.}.

In the characterization of \cite{freistuhler2021nonexistence}, frame {\tt A} is {\em strictly causal}, with 
maximum characteristic speeds less than $1$, while frame {\tt B} is {\em sharply causal} with maximum
characteristic speeds equal to the speed of light (explicit expressions for the characteristic speeds
are given in Sec. \ref{sec:shock_ID_vis}).

Combining (\ref{eq:first_order_transport_coeffs}), (\ref{eq:first_order_constitutive_relations}), and (\ref{eq:decomp_T_ab}) or equivalently (\ref{eq:PF_T_ab})-(\ref{eq:first_order_transport_coeffs}) and (\ref{eq:pi_defn}) gives $T^{ab}$ up to first order; inserting $T^{ab}$ into (\ref{eq:T_ab_conservation_law}) yields the causal, stable relativistic Navier-Stokes equations.

\subsection{Second-order hydrodynamics: M\"uller-Israel-Stewart theory} \label{sec:MIS_covariant}
As mentioned in the introduction, an alternative approach to the unphysical Eckart and Landau-Lifshitz theories was developed by M\"uller \cite{Muller_1967} and Israel and Stewart \cite{Israel_1979} in the 1960's--70's, long before it was known that the choice of hydrodynamic frame was the cause of the pathologies at first order.  In the so-called M\"uller-Israel-Stewart (MIS) formalism, one begins by computing $T^{ab}$ up to second order in gradients of the hydrodynamic variables, at which point one writes the second-order stress-energy tensor as (cf. (\ref{eq:pi_defn}))
\begin{equation}\label{pi2_mis}
T^{ab}_{2} = T^{ab}_{(0)} + \pi^{ab}_{2}.
\end{equation}
The MIS approach differs from that of the BDNK equations, however, in that at first order MIS takes the Landau frame rather than one of the causal, stable frames:
\begin{equation}
\pi^{ab}_{2} = T^{ab}_{(1)} \Big|_{\lambda = \chi = 0} + T^{ab}_{(2)} = \pi^{ab}_{1,L} + T^{ab}_{(2)},
\end{equation}
where we have defined the shorthand
\begin{equation} \label{eq:pi_landau}
\pi^{ab}_{1,L} \equiv T^{ab}_{(1)} \Big|_{\lambda = \chi = 0} = \mathcal{T}^{ab}_{1} = -2 \eta \sigma^{ab}
\end{equation}
for the Landau frame first-order dissipative correction $\pi^{ab}_{1,L}$, which comes from taking the $\lambda = \chi = 0$ case of (\ref{eq:first_order_constitutive_relations})-(\ref{eq:pi_first_order}).

The MIS formalism corrects the pathologies from using the Landau frame by manipulating the second-order terms in the definition of $\pi^{ab}_{2}$.  Writing this definition in compact form, namely showing only $\pi^{ab}_{1,L}$ and one key second-order term while pushing the others into the second-order tensor $\tilde{I}^{ab}_{\pi}$, one has
\begin{eqnarray}\label{MIS_1}
\pi^{ab}_{2} &=& \pi^{ab}_{1,L} + c_0 u^{<c} \nabla_c \sigma^{ab>} + \tilde{I}^{ab}_{\pi} \\
&=& \pi^{ab}_{1,L} - \frac{c_0}{2 \eta} u^{<c} \nabla_c (- 2 \eta \sigma^{ab})^{>} + \frac{c_0}{\eta} u^{<c} \sigma^{ab>} \nabla_c \eta \nonumber \\
& & + \tilde{I}^{ab}_{\pi}. \label{MIS_2}
\end{eqnarray}
Going from equation (\ref{MIS_1}) to (\ref{MIS_2}) above we have replaced $\sigma^{ab}$ in (\ref{MIS_1}) with $-2 \eta \sigma^{ab}$ in (\ref{MIS_2}), adding necessary terms to the latter equation to keep them equal. 
The first step to arrive at the MIS equations is to replace 
$-2 \eta \sigma^{ab} = \pi^{ab}_{1,L}$ (\ref{eq:pi_landau}) with $\pi^{ab}_{2}$ in (\ref{MIS_2}). Recalling
our notation that $\pi^{ab}_{2}=T^{ab}_{(1)}+T^{ab}_{(2)}$, here, since $T^{ab}_{(1)}=-2 \eta \sigma^{ab}$,
this introduces an error that is the gradient of a second-order term, hence is of third order
and negligible. 
Performing the replacement, renaming $\tau_{\pi} \equiv c_0/(2 \eta)$, moving the $\nabla_{c} \eta$ term into a new tensor of second-order terms $I^{ab}_{\pi}$, and rearranging, we find \cite{Baier_2008}
\begin{equation} \label{eq:pi_evolution_eqn}
u^{<c} \, \nabla_c \pi^{ab>}_{2} = \frac{1}{\tau_\pi} (\pi^{ab}_{1, L} - \pi^{ab}_{2}) + I^{ab}_{\pi}.
\end{equation}
This is an advection-type equation
for $\pi^{ab}_2$ with source term that (ignoring $ I^{ab}_{\pi}$) drives the solution toward $\pi^{ab}_{1,L}$ on a timescale determined by the relaxation time transport coefficient $\tau_{\pi}$. The final step in the MIS approach is to now consider $\pi^{ab}_{2}$ as new, independent degrees
of freedom, with (\ref{eq:pi_evolution_eqn}) becoming their evolution equation, and using (\ref{pi2_mis}) 
verbatim in the conservation equation 
(\ref{eq:T_ab_conservation_law}).

The convenience of having another set of evolution equations (\ref{eq:pi_evolution_eqn}) comes at the cost of second-order terms, of which there are a great number.  In (\ref{eq:pi_evolution_eqn}) these terms are hidden in $I^{ab}_{\pi}$, and each acquires a corresponding transport coefficient which must be computed separately using some microscopic theory of the substance being modeled.  Since we are here only interested in first-order dissipative effects on fluid dynamics, we drop $I^{ab}_{\pi}$; this is sometimes called ``truncated'' MIS theory,
though for brevity in Sec. \ref{sec:slab} and beyond will not write ``truncated'' unless 
the distinction is important.  
Dropping $I^{ab}_{\pi}$ violates conformal symmetry \cite{Baier_2008},
so our comparisons between BDNK and MIS evolutions presented
later are more to illustrate how these two theories
provide dissipation in beyond-ideal hydrodynamics,
rather than to serve as a comparison between two models of the same hypothetical
underlying microscopic theory.
Were we to include terms to retain conformal symmetry in MIS,
the two theories would still not be identical at first order
even taking frame transformations into account,
and it is not straightforward to envision how a quantitative ``apples-to-apples''
comparison could be made; we plan to investigate this issue in 
more detail in future work.

Over the nearly sixty years of its existence, a lot has come to be understood about MIS theory, both in general and as it pertains to the study of the QGP.  For a more complete treatment of second-order dissipative hydrodynamics see the review \cite{Romatschke_2019}; for a thorough treatment of conformal second-order terms (BRSSS formalism) see \cite{Baier_2008}; a general discussion of
hyperbolic conformal theories of divergence form can be found in \cite{Lehner:2017yes}; and for a derivation from the Boltzmann equation (DNMR formalism) see \cite{Denicol_2010,Denicol_2012}.  

Much has also been learned about the mathematical properties of the MIS equations of motion, though the added complexity of working at second order has stymied the derivation of some results which are already known for the more recently developed first-order theories.  As was mentioned in section \ref{sec:RNS_covariant}, the BDNK equations are stable, causal, consistent with the second law of thermodynamics, strongly hyperbolic, and well-posed with appropriate constraints on the transport coefficients~\cite{Bemfica_2018,Bemfica_2020}. For MIS on the other hand, the known properties are slightly weaker.  The MIS equations are stable at the linear level, which in turn implies causal propagation \cite{Hiscock_1983}; they are consistent with the second law of thermodynamics by construction \cite{Israel_1979}; they have been shown to be well-posed in the case 
where $\pi^{ab}_{2}$ does not include heat conduction or particle diffusion \cite{Bemfica_2020_MIS}; and they have only been proven to be hyperbolic when all dissipative effects but bulk viscosity are neglected \cite{Bemfica_2019}.  
Nonlinear proofs of stability, causality, local well-posedness, and hyperbolicity do not yet exist for the general case in (3+1)D.

\section{Dissipative fluids in slab-symmetric 4D Minkowski spacetime}\label{sec:slab}

This work is meant to be a first study of the nonlinear dynamics of the BDNK equations, and to compare those solutions with ones obtained using an MIS-based code; to that end, we will focus entirely on the behavior of the fluid and neglect spacetime curvature, specializing to 4D Minkowski spacetime.  Furthermore, to simplify the numerics we will use Cartesian coordinates $x^{a} = (t, x, y, z)^{T}$, and will restrict ourselves to systems which only vary in $t, x$ (``slab'' or ``planar'' symmetry).

In slab-symmetric 4D Minkowski spacetime, the fluid four-velocity may be written
\begin{equation}
u^{a} = (W, W v, 0, 0)^{T},
\end{equation}
where $W \equiv (1-v^2)^{-1/2}$ is the Lorentz factor of the flow.  The two nontrivial hydrodynamic variables are then $\epsilon(t,x)$ and $v(t,x)$, and only the $t, x$ components of (\ref{eq:T_ab_conservation_law}) are nontrivial, $\partial_{c} T^{c t} = 0, \partial_{c} T^{c x} = 0$.  Using the decomposition (\ref{eq:pi_defn}), one may write these equations as 
\begin{align}
0 &= \dot{T}^{tt}_{(0)} + (T^{tx}_{(0)})' + \dot{\pi}^{tt}_{k} + (\pi^{tx}_{k})'  \label{eq:Ttt_evol_eqn} \\
0 &= \dot{T}^{tx}_{(0)} + (T^{xx}_{(0)})' + \dot{\pi}^{tx}_{k} + (\pi^{xx}_{k})' \label{eq:Ttx_evol_eqn}
\end{align}
where the $k = 0$ case corresponds to the perfect fluid equations of motion (relativistic Euler equations), $k = 1$ the BDNK equations, and $k = 2$ the MIS equations.  In the equations above and for the remainder of this work, an overdot represents the time derivative of a quantity $\partial_{t}$, and a prime denotes a spatial derivative $\partial_{x}$.

The following three subsections define the terms in (\ref{eq:Ttt_evol_eqn})-(\ref{eq:Ttx_evol_eqn}), giving the relativistic Euler equations (Sec. \ref{sec:PF_EOM}), BDNK equations (Sec. \ref{sec:RNS_EOM}) and MIS equations (Sec. \ref{sec:MIS_EOM}).

\subsection{Relativistic Euler equations} \label{sec:PF_EOM}

In slab-symmetric 4D Minkowski spacetime, the components of $T^{ab}_{(0)}$ are
\begin{align}
T_{(0)}^{tt} \equiv \tau &= (\epsilon + P) W^2 - P \label{eq:PF_Ttt_eqn}\\
T_{(0)}^{tx} \equiv S    &= v(\tau + P) \label{eq:PF_Ttx_eqn} \\
T_{(0)}^{xx}             &= S v + P, \label{eq:PF_Txx_eqn}
\end{align}
where we have defined the shorthand $\tau, S$ for $T^{tt}_{(0)}, T^{tx}_{(0)}$, respectively, following \cite{Marti_1999,Neilsen_2000,Noble_2003}.  At zeroth order, (\ref{eq:PF_Ttt_eqn})-(\ref{eq:PF_Txx_eqn}) complete the equations of motion (\ref{eq:Ttt_evol_eqn})-(\ref{eq:Ttx_evol_eqn}), as zeroth-order hydrodynamics has no dissipative correction ($\pi^{ab}_{0} = 0$) by definition.  Hence the nontrivial equations of motion for the perfect fluid are
\begin{align}
0 &= \dot{\tau} + S'  \label{eq:PF_t_eqn}\\
0 &= \dot{S} + (S v + P)'. \label{eq:PF_x_eqn}
\end{align}

\subsection{BDNK equations} \label{sec:RNS_EOM}
At first order, the constitutive relations defining $\pi^{ab}_{1}$ take the form
\begin{equation} \label{eq:first_order_const_in_basis}
\begin{aligned}
\mathcal{A}_{1} &= \frac{3}{4} \frac{\chi_0}{\epsilon^{1/4}} W (\dot{\epsilon} + v \epsilon') + \chi_0 \epsilon^{3/4} W^3 (v \dot{v} + v') \\
Q^{x}_{1} &= \frac{\lambda_0}{4 \epsilon^{1/4}} W^2 (v \dot{\epsilon} + \epsilon') + \lambda_0 \epsilon^{3/4} W^4 (\dot{v} + v v') \\
\mathcal{T}^{xx}_{1} &= - \frac{4}{3} \eta_0 \epsilon^{3/4} W^5 (v \dot{v} + v') \\
\end{aligned}
\end{equation}
where the requirement that $\mathcal{Q}^a_{1}$ be transverse implies $Q^{t}_{1} = v Q^{x}_{1}$, and the requirement that $\mathcal{T}^{ab}_{1}$ is transverse and traceless implies $\mathcal{T}^{tt}_{1} = v \mathcal{T}^{tx}_{1} = v \mathcal{T}^{xt}_{1} = v^2 \mathcal{T}^{xx}_{1}$.  Inserting the definitions (\ref{eq:first_order_const_in_basis}) into (\ref{eq:pi_first_order}) gives the components of $\pi^{ab}_{1}$:
\begin{equation}
\begin{aligned}\label{eq:first_order_const_in_basis_b}
\pi^{tt}_1 &= \frac{1}{3} W^2 (3 + v^2) \mathcal{A}_{1} + 2 W v \mathcal{Q}^{x}_{1} + v^2 \mathcal{T}^{xx}_{1} \\
\pi^{tx}_1 &= \frac{4}{3} W^2 v \mathcal{A}_{1} + W (1 + v^2) \mathcal{Q}^{x}_{1} + v \mathcal{T}^{xx}_{1} \\
\pi^{xx}_1 &= \frac{1}{3} W^2 (1 + 3 v^2) \mathcal{A}_{1} + 2 W v \mathcal{Q}^{x}_{1} + \mathcal{T}^{xx}_{1},
\end{aligned}
\end{equation}
which may be combined with the zeroth-order stress-energy tensor components (\ref{eq:PF_Ttt_eqn})-(\ref{eq:PF_Txx_eqn}) to complete the equations of motion (\ref{eq:Ttt_evol_eqn})-(\ref{eq:Ttx_evol_eqn}), yielding
\begin{equation} \label{eq:BDNK_eqns}
\begin{aligned}
0 &= \dot{\tau} + S' + \dot{\pi}^{tt}_{1} + (\pi^{tx}_{1})' \\
0 &= \dot{S} + (S v + P)' + \dot{\pi}^{tx}_{1} + (\pi^{xx}_{1})'. \\
\end{aligned}
\end{equation}

\subsection{MIS equations} \label{sec:MIS_EOM}
Since $\pi^{ab}_{2}$ is defined to be symmetric, transverse to $u^{a}$, and traceless, we have the identities \cite{Baier_2008}
\begin{equation}\label{mis_ident}
\pi^{tt}_{2} = v \pi^{tx}_{2} = v \pi^{xt}_{2} = v^2 \pi^{xx}_{2}.
\end{equation}  
As a result, evolving $\pi^{xx}_{2}$ is sufficient to constrain the whole tensor\footnote{Only $\pi^{xx}_{2}$ is needed as long as $\pi^{yy}_{2}, \pi^{zz}_{2}$ are initialized to zero, as is the case here.} $\pi^{ab}_{2}$, and we will only need the $xx$ component of (\ref{eq:pi_evolution_eqn}), which is
\begin{multline} \label{eq:pi_evol_eqn_in_basis}
\dot{\pi}^{xx}_{2} + v (\pi^{xx}_{2})' = \frac{1}{W \tau_{\pi}} (\pi^{xx}_{1,L} - \pi^{xx}_{2}) \\
+ 2 W^2 v \pi^{xx}_{2} \dot{v} + 2 W^2 v^2 \pi^{xx}_{2} v',
\end{multline}
where the Landau frame first-order dissipative correction is $\pi^{xx}_{1,L} = \mathcal{T}^{xx}_{1}$ from (\ref{eq:first_order_const_in_basis}).  The equations of motion for the MIS system are then (\ref{eq:pi_evol_eqn_in_basis}) to evolve $\pi^{xx}_{2}$, and the two nontrivial components of the stress-energy conservation equation:
\begin{align}
0 &= \dot{\tau} + S' + \dot{\pi}^{tt}_{2} + (\pi^{tx}_{2})' \\
0 &= \dot{S} + (S v + P)' + \dot{\pi}^{tx}_{2} + (\pi^{xx}_{2})'.
\end{align}

\section{Numerical Methods} \label{sec:numerical_method}

\subsection{Conservative schemes for ideal hydrodynamics} \label{sec:conserv_schemes_ideal_hydro}

The ultimate goal when writing down a fluid model is to determine the time evolution of the hydrodynamic variables.  With this in mind, a naive way to formulate a numerical method to solve (\ref{eq:T_ab_conservation_law}) at zeroth order in gradients (ideal hydrodynamics) would be to treat it as a set of evolution equations for $\epsilon, v$ explicitly, e.g. the $t, x$ components of $\nabla_{a} T^{ab}_{0} = 0$, (\ref{eq:PF_t_eqn})-(\ref{eq:PF_x_eqn}), would be written
\begin{equation} \label{eq:naive_scheme}
\begin{aligned}
\dot{\epsilon} &= F(\dot{v}, \epsilon', v', \epsilon, v) \\
\dot{v} &= G(\dot{\epsilon}, \epsilon', v', \epsilon, v),
\end{aligned}
\end{equation}
for some nonlinear functions $F, G$.  One would then solve a discretization of the coupled nonlinear PDEs (\ref{eq:naive_scheme}) to evolve $\epsilon, v$ forward in time.

A naive scheme of the form (\ref{eq:naive_scheme}) should work in principle as long as the solutions are smooth.  However, solutions to the relativistic Euler equations (\ref{eq:T_ab_conservation_law}), (\ref{eq:PF_Ttt_eqn})-(\ref{eq:PF_Txx_eqn}), are not generically smooth, as discontinuities in $\epsilon, v$ (\textit{shockwaves}) can form dynamically \cite{Pan_2005,Gremaud_2014}.  In these cases the physical solution is given not by direct solution of the PDEs (\ref{eq:naive_scheme})---as derivative terms $\epsilon', v'$ diverge---but instead by solution to the weak formulation of the equations \cite{Smoller_1993}.

To resolve shocks in ideal hydrodynamics, instead of (\ref{eq:naive_scheme}) one writes 
(\ref{eq:PF_Ttt_eqn}-\ref{eq:PF_Txx_eqn}) in so called {\em flux conservative form}
\begin{equation} \label{eq:general_conservation_law}
\frac{\partial}{\partial t} \bm{q} + \frac{\partial}{\partial x^{i}} \bm{f}_{[i]} = \bm{\psi},
\end{equation}
where the vector $\bm{q}$ is populated with \textit{conservative variables}, $\bm{f}_{[i]}$ is the $i$th component of a vector of \textit{fluxes} (with $i$ restricted to spatial indices), $\bm{\psi}$ is a vector of \textit{sources}, and each is a function of the \textit{primitive variables} $\bm{p}$ (in this case, $\bm{p} = (\epsilon, v)^{T}$).  This approach is specialized to conservation laws, and allows one to apply special methods rooted in the weak formulation of the equations to handle the spatial derivative term, $\partial\bm{f}_{[i]}/\partial x^{i}$, when discontinuities are present.  Among these methods are artificial viscosity techniques, which smooth shocks until they no longer destabilize the numerical scheme, and high-resolution shock-capturing (HRSC) methods, which use the characteristic structure across a discontinuity to derive a discretization for $\bm{f}$ that is stable across it.  For a detailed summary of these methods, see for example the reviews of Mart\'i and M\"uller \cite{Marti_1999}, Font \cite{Font_2000}, and LeVeque's book \cite{LeVeque_2006}.

Note that it is typically unfeasible and sometimes impossible to analytically solve for the primitive variables $\bm{p}$ as explicit functions of the conservative variables $\bm{q}$; hence the flux $\bm{f}_{[i]}$ and the source term $\bm{\psi}$ are generically written as functions of both $\bm{q}$ and $\bm{p}$.  Since a solution to (\ref{eq:general_conservation_law}) only provides updated values of $\bm{q}$, it becomes necessary to compute $\bm{p}$ from the updated variables $\bm{q}$ in order to perform the next time evolution step.  This process of computing $\bm{p}(\bm{q})$, sometimes called \textit{primitive variable recovery}, often involves solving a system of coupled nonlinear algebraic equations and occurs many times within a time step.  For this reason it is often the most time consuming part of the numerical scheme; fortunately a number of algorithms have been discovered for the standard sets of conservative and primitive variables, and the computational cost is usually not prohibitive \cite{Marti_1999}.

\subsection{Conservative formulations for the relativistic Euler, BDNK, and MIS equations}
In this subsection we will cast the zeroth-order relativistic Euler, first-order BDNK, and second-order MIS equations into conservative form (\ref{eq:general_conservation_law}). 

\subsubsection{Zeroth order: relativistic Euler equations}
Starting at zeroth order, comparing the relativistic Euler equations (\ref{eq:PF_t_eqn})-(\ref{eq:PF_x_eqn}) with (\ref{eq:general_conservation_law}), we can see that
\begin{equation} \label{eq:PF_conserv_terms}
\bm{q}^{PF} = 
\begin{pmatrix}
\tau \\ 
S
\end{pmatrix}, ~~~
\bm{f}^{PF} = 
\begin{pmatrix}
S \\
S v + P
\end{pmatrix}, ~~~
\bm{\psi}^{PF} = \bm{0}.
\end{equation}
It turns out that the high degree of symmetry in the conformal fluid $T^{ab}_{0}$ allows one to do the primitive variable recovery analytically, and one finds $\bm{p}(\bm{q})$ to be \cite{Neilsen_2000}
\begin{equation} \label{eq:PF_prim_var_recovery}
\begin{aligned}
\epsilon &= - \tau + \sqrt{4 \tau^2 - 3 S^2} \\
v &= \frac{3 S}{3 \tau + \epsilon}.
\end{aligned}
\end{equation}

\subsubsection{First order: BDNK equations}\label{BDNK1}

Since $T^{ab}_{1}$ is first order in gradients, 
the BDNK equations (\ref{eq:Ttt_evol_eqn})-(\ref{eq:Ttx_evol_eqn}) are second order PDEs. Hence, as mentioned in the introduction, one
would expect to have to adapt numerical methods to this structure, rather than being able to use methods devised for the Euler equations (which contain only first derivatives).
If one wants to keep the equations in conservation-law form,
one can do so by performing a first order reduction in time, and instead taking the primitive variables to be $\bm{p}^{NS} \equiv (\dot{\epsilon}, \dot{v})^{T}$.  The BDNK equations then take the form (\ref{eq:general_conservation_law}) 
with
\begin{equation} \label{eq:RNS_conserv_form}
\bm{q}^{NS} =
\begin{pmatrix}
\pi^{tt}_{1} \\
\pi^{tx}_{1}
\end{pmatrix}, ~~~
\bm{f}^{NS} = \bm{f}^{PF} + \bm{f}^{\pi}_{1}, ~~~
\bm{\psi}^{NS} = -\dot{\bm{q}}^{PF},
\end{equation}
where
\begin{equation} \label{eq:dissipative_flux}
\bm{f}^{\pi}_{k} \equiv 
\begin{pmatrix}
\pi^{tx}_{k} \\
\pi^{xx}_{k}
\end{pmatrix}
\end{equation}
which appears with $k = 1$ in (\ref{eq:RNS_conserv_form}).
For simplicity we do not include in the conservative system the ``trivial'' evolution equations $d\epsilon/dt=\dot{\epsilon}$ and $dv/dt=\dot{v}$ that are used to update {$\epsilon,v$} (if one did, $(\epsilon,v)$ would be added to
the vector of conservative variables, and their corresponding flux and source terms
would be $(0,0)$ and $(\dot{\epsilon},\dot{v})$ respectively).

For the BDNK system, the conservative variables are linear functions of the primitive variables,
and it is straightforward 
to solve for $\bm{p}^{NS}(\bm{q}^{NS})$ analytically; the results are lengthy and not particularly
illuminating, so we list them in Appendix \ref{sec:RNS_prim_var_recovery}.

\subsubsection{Second order: MIS equations}

For the MIS formalism, one is able to use the additional evolution equation for $\pi^{xx}_{2}$ (\ref{eq:pi_evol_eqn_in_basis}) to evolve all of the first and second-order terms from $T^{ab}_{2}$.  
It is not a conservation law, and may be solved using standard methods.

The presence of (\ref{eq:pi_evol_eqn_in_basis}) allows us to use (\ref{eq:Ttt_evol_eqn})-(\ref{eq:Ttx_evol_eqn}) to evolve $\tau, S$ as in the perfect fluid case, and accordingly allows us to cast (\ref{eq:Ttt_evol_eqn})-(\ref{eq:Ttx_evol_eqn}) in conservative form (\ref{eq:general_conservation_law}) with the same set of conservative variables, hence the same $\bm{p}(\bm{q})$
primitive variable recovery scheme (\ref{eq:PF_prim_var_recovery}). The full set of terms are
\begin{equation}\label{mis_eqns}
\bm{q}^{MIS} = \bm{q}^{PF}, ~~~ \bm{f}^{MIS} = \bm{f}^{PF} + \bm{f}^{\pi}_{2}, ~~~ \bm{\psi}^{MIS} = -
\begin{pmatrix}
\dot{\pi}^{tt}_{2} \\
\dot{\pi}^{tx}_{2}
\end{pmatrix},
\end{equation}
with $\bm{f}^{\pi}_{2}=({\pi}^{tt}_{2},{\pi}^{tx}_{2})^{T}$, (\ref{eq:dissipative_flux}).

\subsection{Discretization}
We use a finite volume approach to discretize the fluid equations of motion, dividing the domain into cells of area $\Delta x \Delta t$ bounded by $[x_{i-1/2}, x_{i+1/2}]$ in space and $[t^{n}, t^{n+1}]$ in time.  Continuum fields describing the fluid $C$ are then replaced with their cell averages $C^{n}_{i}$.  For all of the simulations performed here, we divide the spatial domain into $N$ cells, with $N-1 = 2^{7}$ to $2^{12}$, and use a Courant factor $\lambda \equiv \Delta t/\Delta x = 0.1$.  For the smooth Gaussian test problem we use a periodic domain (identifying cell $0$ with cell $N-1$), and no boundary conditions are needed. For the other two tests, at the spatial boundaries of the domain, the outermost two cells at each end ($i = 0, 1, N-2, N-1$) are designated ghost cells, whose values are not evolved using the discretized PDEs, but are instead copied from the nearest non-ghost cell. Explicitly, at time level $n$ we copy the value $C^{n}_{2}$ into $C^{n}_{0}, C^{n}_{1}$, and $C^{n}_{N-3}$ into $C^{n}_{N-2}, C^{n}_{N-1}$. Convergence tests are described in Appendix \ref{sec:convergence_tests}.

\subsubsection{Zeroth order: relativistic Euler equations} \label{sec:PF_discretization}
We discretize the relativistic Euler equations using the method of lines, following \cite{Neilsen_2000,Noble_2003}. Specifically, 
we evolve in time using Heun's method (an explicit second-order Runge-Kutta-type scheme) \cite{Noble_2003,LeVeque_2006,Schenke_2012}. Writing (\ref{eq:general_conservation_law}) as $\dot{\bm{q}} = \bm{\psi} - \bm{f}' \equiv \bm{H}(\bm{q})$, Heun's method updates $\bm{q}$ in two steps via
\begin{equation} \label{eq:Heuns_method}
\begin{aligned}
\bar{\bm{q}}^{n+1} &= \bm{q}^{n} + \Delta t \, \bm{H}(\bm{q}^{n}) \\
\bm{q}^{n+1} &= \bm{q}^{n} + \frac{\Delta t}{2} \Big[ \bm{H}(\bm{q}^{n}) + \bm{H}(\bar{\bm{q}}^{n+1}) \Big].
\end{aligned}
\end{equation}

To discretize the flux term $\bm{f}'$, we use the Roe approximate Riemann solver \cite{Roe_1997} along with the minmod slope limiter \cite{Neilsen_2000}.

\subsubsection{First order: BDNK equations} \label{sec:RNS_discretization}
Shock-capturing methods were developed for the relativistic Euler system because the equations are known to possess physical, discontinuous shock solutions \cite{Smoller_1993}. As discussed in the introduction, it is unclear whether solutions with discontinuities in the hydrodynamic variables can be made sense of for the BDNK or MIS equations.  Even if such solutions are mathematically sensible, their infinite gradients would make them untrustworthy from the perspective of the gradient expansion. However, since we are ultimately interested in applications where sharp transitions may develop over scales too small to resolve, it would behoove us to use methods that can deal with such effective discontinuities. With that in mind, we use a simple scheme that is able to evolve the kind of discontinuous initial data used in our shock tube test, at least if the discontinuity and/or viscosity is not too large.

For large jumps or large viscosity---the region of parameter space where the gradient expansion should break down---our numerical method fails\footnote{Our algorithm also breaks down in the typical problematic regimes experienced by many relativistic hydrodynamic codes, e.g., flow velocities approaching the speed of light, or very low densities. To help distinguish those failures from ones that may be associated with viscosity, one can monitor the series of tests (see Sec. \ref{sec:diagnostics}) designed to indicate whether one is evolving outside of the regime of validity of the gradient expansion.}.
In addition, since we have based our algorithm on a conservative form of the equations adapted to their principal structure (Sec. \ref{BDNK1}), 
it does not work with exactly zero viscosity (and in practice neither for viscosity so small that the primitive variable
recovery (\ref{eq:RNS_epsdot})-(\ref{eq:RNS_vdot}) becomes dominated by round-off error, 
as those expressions have a $0/0$ form in the limit
$\eta\rightarrow 0$). If being able to run with exactly zero viscosity is important for a BDNK scheme, then a different
set of variables and solution algorithm would be required (for example, one more akin to that used for the MIS equations
described in the following subsection).

As with the Euler equations (Sec. \ref{sec:PF_discretization}), we use Heun's method to evolve $\bm{q}$ forward in time.
We also compute the perfect fluid contribution $\bm{f}^{PF}$ to the flux term $\bm{f}^{NS} = \bm{f}^{PF} + \bm{f}^{\pi}_{1}$
as with the Euler equations, namely using the Roe flux with minmod limiter.
The main difference for the BDNK equations then is how we deal with the viscous part $\bm{f}^{\pi}_{1}$ of the flux.
For this, we effectively treat it as a source term, discretizing the spatial derivative $(\bm{f}^{\pi}_{1})'$ using standard centered, 
second-order-accurate finite difference stencils\footnote{For smooth initial data, 
even the perfect fluid flux can be computed with finite differences---using the Roe flux is only necessary at early times
for the shock tube test.}.
Regarding that, it is crucial to note that of the two components of the flux $\bm{f}^{\pi}_{1} = (\pi^{tx}_{1}, \pi^{xx}_{1})^T$, only the first is a conservative variable $\bm{q}$ being dynamically evolved. For the second component $\pi^{xx}_{1}$ then, we need to replace it with its definition
(\ref{eq:first_order_constitutive_relations},\ref{eq:pi_first_order}), which contains derivative terms such as $\epsilon', v'$. Thus,
the gradient of the corresponding flux term contains second spatial derivatives, that we also discretize using a standard centered second-order-accurate finite difference stencil.  Note that we do not need to use mixed space-time difference operators, as our primitive variables are $\bm{p} = (\dot{\epsilon}, \dot{v})^{T}$, i.e. in the gradient of the flux term it is simply their spatial derivatives that appear.
For the sake of clarity, we provide a detailed list of the actions performed during one time evolution step of our
BDNK numerical algorithm in Appendix \ref{sec:RNS_pseudocode}.

When evolving discontinuous initial data, at early times we find adding Kreiss-Oliger style dissipation \cite{Kreiss_1973} helps
in achieving stable evolution. Specifically, during both the predictor and corrector step of the time integration we apply this artificial 
dissipation to $\pi^{tx}_1$ and $\pi^{xx}_1$ with amplitude coefficient $\alpha_{KO} \sim 0.1$. Kreiss-Oliger dissipation is unnecessary for evolutions starting from smooth initial data, and ceases to be necessary
shortly after physical dissipation smooths the shock in cases with discontinuous initial data.

\subsubsection{Second order: MIS equations}
For the MIS equations, we discretize the $\pi^{xx}_{2}$ evolution equation (\ref{eq:pi_evol_eqn_in_basis}) using a simple first-order upwind scheme \cite{Okamoto_2017,Takamoto_2011}. Explicitly, we write the advection operator as
\begin{equation} \label{eq:upwind_discretization}
(\partial_{t} + v \partial_{x}) C \approx \dot{C} +
\begin{cases}
v^{n}_{i} \frac{C^{n}_{i} - C^{n}_{i-1}}{\Delta x} & v^{n}_{i} \geq 0 \\
v^{n}_{i} \frac{C^{n}_{i+1} - C^{n}_{i}}{\Delta x} & v^{n}_{i} < 0, \\
\end{cases}
\end{equation}
where the time evolution of $\dot{C}$ is again performed using Heun's method, and all remaining spatial derivatives outside the advection operator (such as $v'$) are handled with centered, second-order-accurate finite differences.

For the conservation law (\ref{eq:general_conservation_law}) we again use Heun's method for the time evolution.  We also follow the BDNK approach by splitting the flux into a perfect fluid piece and a dissipative piece, using a Roe solver and finite differences for $(\bm{f}^{PF})'$ and $(\bm{f}^{\pi}_{2})'$, respectively. MIS differs from BDNK though in that $(\bm{f}^{\pi}_{2})'$ only requires first differences of $v$ and $\pi^{xx}_2$ (\ref{mis_ident}).

We handle the source term $\bm{\psi}^{MIS}$ (\ref{mis_eqns}) in the same way as \cite{Schenke_2012}, using a backward time difference $\dot{C} \approx (C^{n}_{i}-C^{n-1}_{i})/\Delta t$ in the predictor step of Heun's method (computing $\bar{\bm{q}}$ in (\ref{eq:Heuns_method})).  In the corrector step we use the advanced time level from the predictor step, $\dot{C} \approx (\bar{C}^{n+1}_{i}-C^{n}_{i})/\Delta t$.

\section{Physical regime of interest}\label{sec:phys}
We adopt natural units, which means that a quantity with SI units $\textnormal{kg}^{\alpha} \textnormal{m}^{\beta} \textnormal{s}^{\gamma}$ is written $(E)^{\alpha-\beta-\gamma} \hbar^{\beta+\gamma} c^{\beta-2\alpha}$, where $E$ is an energy unit (e.g. GeV, J, etc.) and the factors of $\hbar, c$ may be ignored once one sets the fundamental constants $c = \hbar = k_B = 1$.  As a result of conformal symmetry, the choice of energy unit in this case fixes an overall energy scale, but does not meaningfully alter the dynamics\footnote{For example, consider the effect of a change in units $E \to E' = \lambda E$ on the spacetime evolution of the system, $T^{ab}(x^{a})$.  The transformation takes $T^{ab} \to \lambda^{4} T^{ab}$ and $x^{a} \to \lambda^{-1} x^{a}$; the latter is a symmetry of the stress-energy tensor due to conformal invariance, and hence the net effect of $E \to E'$ is just the constant rescaling $T^{ab}(x^{a}) \to \lambda^{4} T^{ab}(x^{a})$.  Another way to see this is to notice that $\lambda$ cancels from the equations of motion, (\ref{eq:T_ab_conservation_law}).}. With this in mind, for the remainder of this work we (arbitrarily) choose to measure energies in GeV.

We derive intuition from the phenomenology of heavy-ion collisions to make our choice for $\eta_0$, which determines the amount of viscosity in the solution.
QGP viscosities have been measured to be within 
about 10\% of the so-called KSS bound \cite{Bernhard:2019bmu}
which gives the predicted minimum ratio of shear viscosity $\eta$ to entropy density $s$ for any fluid \cite{Kovtun_2005}:
\begin{equation}
\frac{\eta^{min}}{s} = \frac{1}{4 \pi}.
\end{equation}
We can compute the value of our free parameter $\eta_0$ required to reach the KSS bound using (\ref{eq:first_order_transport_coeffs}) and the entropy density for a conformal fluid \cite{Bemfica_2018}
\begin{equation*}
s = \frac{\epsilon + P}{T},
\end{equation*}
giving
\begin{equation*}
\eta^{min}_0 = \frac{\epsilon_0^{1/4}}{3 \pi}.
\end{equation*}
For a QGP the ratio $\epsilon/T^4 = \epsilon_0 \sim 10$ \cite{Kumar_2018,Snellings_2003}.
In our results below, we consider fluids with $\epsilon_0=10$, and viscosities ranging between 
the KSS bound $\eta/s = (4 \pi)^{-1}$ and $\eta/s = 20 \cdot (4 \pi)^{-1}$.

Despite nearly saturating the KSS lower bound for entropy-normalized viscosity $\eta/s$, the QGP has a high viscosity $\eta$ by everyday standards, which is then compensated for by a correspondingly high entropy density $s$.  In SI units, the QGP viscosity is roughly $\eta \sim 10^{12} \textnormal{ Pa} \cdot \textnormal{s}$ \cite{McInnes_2017}, nearly $10^{13}$ times that for water at STP, despite the fact that water's entropy-normalized viscosity is many times larger, $\eta/s \sim 380 \cdot (4 \pi)^{-1}$ \cite{Kovtun_2005}.

The convergence of the gradient expansion is determined entirely by the size of gradients (such as $\sigma^{ab}$) and transport coefficients (such as $\eta$), not by normalized quantities like $\eta/s$.  Since the QGP has a large shear viscosity $\eta$ and has variation on scales of order fm, both the transport coefficients and gradients in the expansion (\ref{eq:gradient_expansion}) are relatively large.  Hence, if the BDNK and MIS equations can accurately model the dynamics of the QGP, it would be reasonable to expect similar success in regimes where the transport coefficients and gradients are smaller, as is often the case in astrophysics.  One example would be oscillations in an isolated, cold neutron star: neutron star cores are predicted to have viscosities a factor of $\sim 10^{6}$ times larger \cite{Shternin_2008} than the QGP, but variation on scales of km, making gradients at least a factor of $\sim 10^{18}$ smaller. 
Following a binary neutron star merger~\cite{Duez:2018jaf}, if the remnant does not promptly collapse to a black hole, a differentially rotating star would form with much smaller lengthscale variations and higher temperatures (due to shock heating from the collision,
reaching $\sim 10$ MeV, which compares to $\sim 150$ MeV for the QGP \cite{Duez:2018jaf,Alford_2018,Gazdzicki_2016}). However, these 
conditions are likely still well within the regime of validity of the 
BDNK and MIS equations (unlike the QGP-inspired examples we show below, where already at 20 times the KSS bound we see, for example,
violations of the weak energy condition in BDNK evolutions).

For the MIS system, in addition to the viscosity $\eta$ we have another degree of freedom: the relaxation time $\tau_{\pi}$.  Using holographic arguments, \cite{Baier_2008} finds it to be 
\begin{equation} \label{eq:holographic_relaxation_time}
\tau_{\pi} = \frac{2 - \ln 2}{2 \pi T} = \frac{(2 - \ln 2) \epsilon_0^{1/4}}{2 \pi \epsilon^{1/4}}.
\end{equation}
For the sake of simplicity, we follow \cite{Takamoto_2011,Okamoto_2017} in setting it to be a constant\footnote{Though it is not an issue for our purposes, it is important to note that choosing $\tau_{\pi}$ to be constant violates conformal symmetry \cite{Baier_2008} --- see Sec. \ref{sec:MIS_covariant}.},
specifically $\tau_{\pi} = 0.3 ~\textnormal{GeV}^{-1}$ unless otherwise stated.
The chosen value is somewhat smaller than if we were to use (\ref{eq:holographic_relaxation_time}) for the Gaussian and shock tube test we show below, which have a maximum energy density $\epsilon = 0.4 ~\textnormal{GeV}^{4}$ (implying $\tau_{\pi} \approx 0.47 ~\textnormal{GeV}^{-1}$). The shockwave test has $\epsilon$ larger by a factor of a few. 
On the other hand, here
we are actually not interested in treating $\tau_{\pi}$ as an additional, physical transport coefficient; rather, 
it is a device to drive the independent tensor $\pi^{ab}_{2}$ toward the
first-order dissipative tensor $\pi^{ab}_{1, L}$ (see (\ref{eq:pi_evolution_eqn}) with $I^{ab}_{\pi}=0$) that contains
the physics we are interested in modeling. Thus, we want $\tau_{\pi}$ to be small enough that it does not
affect the results, but not so small as to require prohibitively small time steps for stable numerical evolution;
$\tau_{\pi} = 0.3 ~\textnormal{GeV}^{-1}$ is a good choice in that regard. Varying $\tau_{\pi}$ by factor
of a few causes negligible differences in the results for most of the cases studied below,
the exception being in far from equilibrium
scenarios, where for the sake of illustration we also present an example with $\tau_{\pi} = 30 ~\textnormal{GeV}^{-1}$.

\section{Monitoring convergence of the gradient expansion} \label{sec:diagnostics}
The BDNK and MIS theories described here are only well justified modeling dissipative hydrodynamics in regimes where the gradient expansion (\ref{eq:gradient_expansion}) converges.  
Though we are unable to make claims about the convergence or divergence of the gradient series for the nonlinear numerical solutions presented here\footnote{Such claims can be made for highly symmetric flows --- see \cite{Heller:2013fn,Buchel:2016cbj,Denicol:2016bjh,Heller:2016rtz,Grozdanov:2019kge}.}, 
one expects that a truncation at order $k+1$ should be reliable when its contribution to the stress-energy tensor is smaller than the contribution at order $k$.  As such we compute the quantity $|T^{tt}_{(1)}/T^{tt}_{(0)}|$ for the BDNK and MIS solutions, taking $T^{tt}_{(1)} = \pi^{tt}_{1}$ for the former and $T^{tt}_{(1)} = \pi^{tt}_{1, L} = v^2 \pi^{xx}_{1,L}$ for the latter. In regions where $|T^{tt}_{(1)}/T^{tt}_{(0)}| \gtrsim 1$, one would expect higher order terms to be important, and the first-order results to no longer be trustworthy.

The authors of \cite{Bemfica_2020} also suggest checking that the weak energy condition remains satisfied, namely $X_{a} X_{b} T^{ab} \geq 0 \, \forall \, X^{a}$ with $X_{c} X^{c} = -1$, as its violation may indicate entry into a regime in which (\ref{eq:gradient_expansion}) no longer converges.  Along these lines we monitor two choices for $X^a$: the fluid four velocity $u^{a}$
and the simulation reference frame four velocity $(\partial/ \partial t)^a$.
For the BDNK system we also check if $|\mathcal{A}_1/\epsilon|$, (\ref{eq:first_order_constitutive_relations})-(\ref{eq:pi_first_order}), approaches or exceeds unity.

\section{Results} \label{sec:results}

In this section we discuss numerical solutions to the relativistic Euler, BDNK, and MIS equations for three distinct sets of initial data: (A) a smooth, initially stationary profile, (B) a discontinuous (shock tube) setup, and (C) a smooth transition from a supersonic flow at the left boundary to subsonic flow at the right boundary. 
In all cases, this amounts to particular choices of $\epsilon(t=0,x)$ and $v(t=0,x)$. 
For the Euler equations, that completes specification
of the initial data. 

For the BDNK equations, we additionally need to specify $\pi^{tt}_1(t=0,x)$ and $\pi^{tx}_1(t=0,x)$
(or equivalently $\dot{\epsilon}(t=0,x)$ and $\dot{v}(t=0,x)$ from (\ref{eq:RNS_epsdot})),
and for the MIS equations $\pi^{xx}_2(t=0,x)$. In all cases for BDNK we set $\pi^{tt}(t=0) = \pi^{tx}(t=0) = 0$, and for MIS $\pi^{xx}(t=0) = 0$. 
For the MIS equations this always results in the initial evolution being identical
to the perfect fluid at $t=0$. For the BDNK equations this will only be so
if $v(t=0,x)=0$, as is the case for tests (A) and (B), though not so for the shockwave test (C) 
(if desired one can always choose $\dot{\epsilon}(t=0,x)$ and $\dot{v}(t=0,x)$ to be
equal to that of the perfect fluid, but for (C) we are
more interested in understanding the nature of strong shock solutions within BDNK
than comparing to the perfect fluid evolution).

\subsection{Smooth, stationary initial data}
\begin{figure}
	\centering
	\includegraphics[width=\columnwidth]{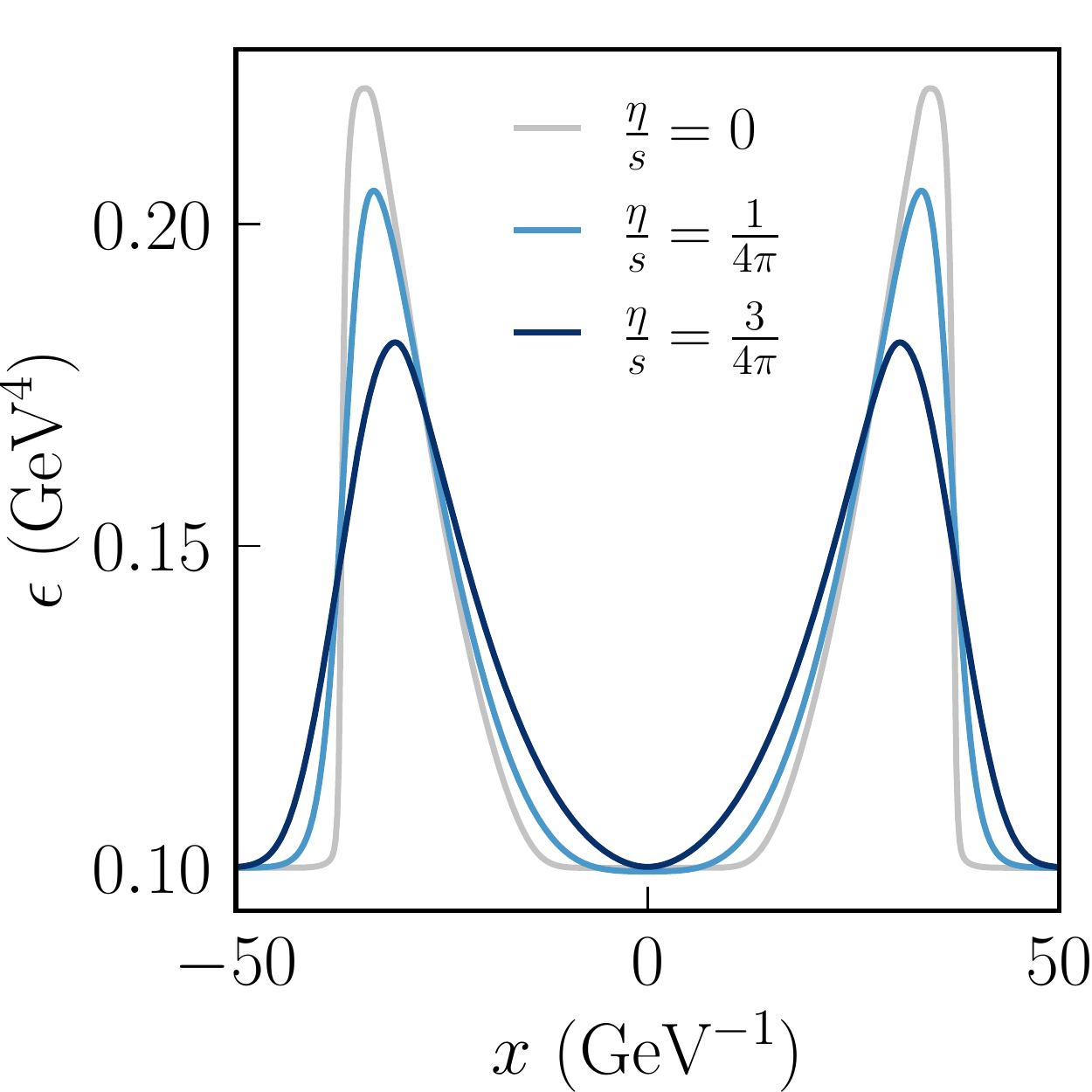}
	\caption{Qualitative effect of viscosity on the evolution of smooth initial data for $\eta/s = \{0, 1, 3\} \cdot (4 \pi)^{-1}$ at $t = 47 ~\textnormal{GeV}^{-1}$, computed using the relativistic Euler equations (\ref{eq:PF_t_eqn})-(\ref{eq:PF_x_eqn}) for $\eta=0$, and the BDNK equations (\ref{eq:first_order_const_in_basis})-(\ref{eq:BDNK_eqns}) for the viscous cases (MIS solutions for these would appear identical---see Fig.~\ref{fig:FOCS_MIS_comp_smooth}).  As expected, viscosity smooths the profile in $\epsilon$ compared to the perfect fluid case. Also evident is the steepening of the leading feature of each pulse, which for the perfect fluid case forms a step function discontinuity at $x \approx \pm 38 ~\textnormal{GeV}^{-1}$ (discontinuities do not form for the two viscous cases).} \label{fig:smooth_ID_qualitative}
\end{figure}

We first consider the evolution of data that is initially stationary $v(t = 0, x) = 0$, and has a smooth Gaussian profile in the energy density
\begin{equation} \label{eq:gaussian_ID}
\epsilon(t=0, x) = A e^{-x^2/w^2} + \delta.
\end{equation}
For a concrete example we choose the amplitude $A = 0.4 ~\textnormal{GeV}^{4}$, width $w = 25 ~\textnormal{GeV}^{-1}$, and background energy density $\delta = 0.1 ~\textnormal{GeV}^{4}$. 
Fig. \ref{fig:smooth_ID_qualitative} shows a snapshot of $\epsilon(t,x)$ at $t = 47 ~\textnormal{GeV}^{-1}$, run with three values of the viscosity $\eta/s = \{0, 1, 3\} \cdot (4 \pi)^{-1}$.
The viscous evolutions in this figure were produced with the BDNK equations using frame {\tt A} (\ref{frames}), 
but look identical (at the scale of the figure)
to the corresponding cases evolved with the MIS equations.
By the time shown in the figure, 
the initial Gaussian profile in $\epsilon$ has split into two clumps that are propagating away from each other.
One can clearly see from the figure that viscosity acts to smooth sharp features in the energy density profile (and similarly in the 
velocity profiles that develop).  Despite the fact that the flow velocities are initialized to zero, the outer edges of the perfect fluid profile dynamically become supersonic, and a step function discontinuity can be seen at $x \approx \pm 38 ~\textnormal{GeV}^{-1}$; discontinuities do not form in the viscous cases with this initial data.

\begin{figure*}
	\centering
	\includegraphics[width=\textwidth]{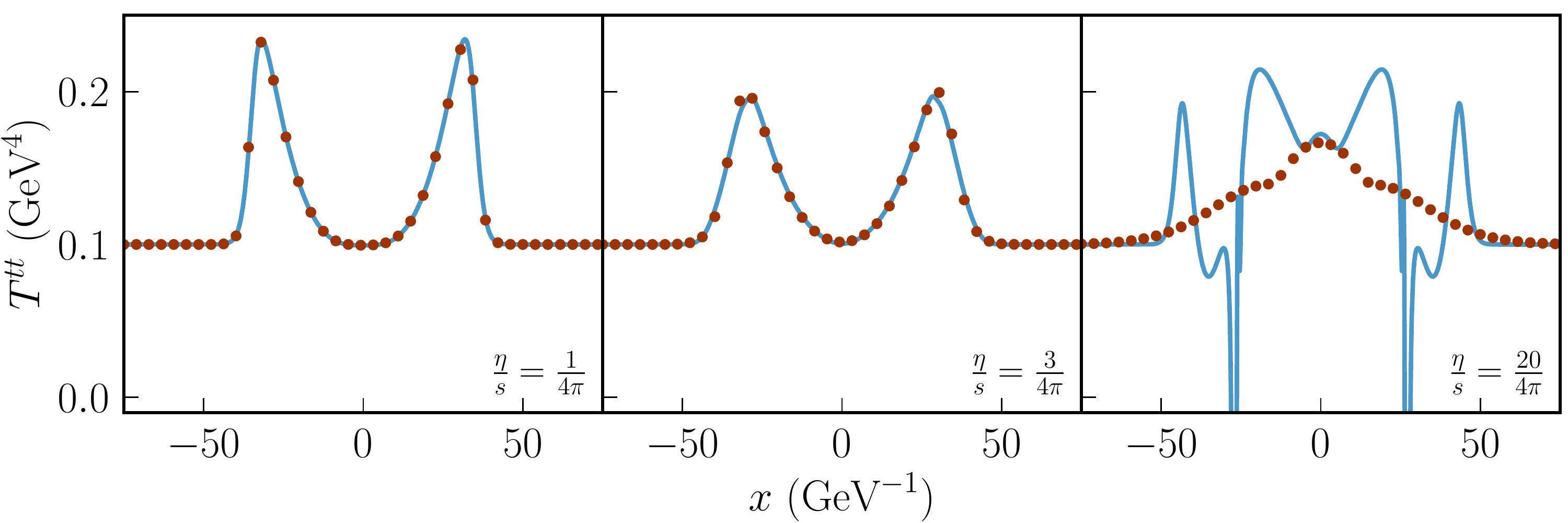}
	\caption{Comparison of solutions for the $tt$ component of the stress-energy tensor, $T^{tt}$, obtained with the BDNK theory (frame {\tt A} (\ref{frames})) and MIS theory, in lines and dots respectively, at $t = 35 ~\textnormal{GeV}^{-1}$ for viscosities $\eta/s = \{1, 3, 20\} \cdot (4 \pi)^{-1}$ from left to right.  Note that at the two lower viscosities, the solutions are qualitatively identical for the BDNK and MIS equations.  In the highest viscosity case (rightmost panel) BDNK theory gives a qualitatively different solution from MIS and the lower viscosity solutions, instead forming multiple maxima, developing sharp features, and even changing sign (the MIS solution shown is in the process of splitting into two clumps, as in the lower-viscosity cases). There is evidence that this solution lies outside of the regime of validity of the gradient expansion at first order---see Fig. \ref{fig:FOCS_MIS_diagnostics}. Note that to avoid clutter the MIS points are a sparse sampling of the actual resolution of the simulation. } \label{fig:FOCS_MIS_comp_smooth}
\end{figure*}

In Fig.~\ref{fig:FOCS_MIS_comp_smooth} we compare solutions of the BDNK (frame {\tt A}) and MIS equations (blue lines and red dots, respectively) with a sufficiently large viscosity $\eta/s = 20 \cdot (4 \pi)^{-1}$ (right panel) that they show markedly different evolution 
(for reference, in the left and center panels we also show the two viscous cases from Fig.~\ref{fig:smooth_ID_qualitative}, though
this snapshot is at a slightly earlier time). With time, the BDNK case splits into four clumps in $\epsilon$ rather than two.  The MIS solution still splits into two clumps, though at the time shown in Fig.~\ref{fig:FOCS_MIS_comp_smooth} it is in the midst of doing so; it eventually settles to a state qualitatively similar to the lower viscosity cases shown in
the left and center panels.

\begin{figure}
	\centering
	\includegraphics[width=\columnwidth]{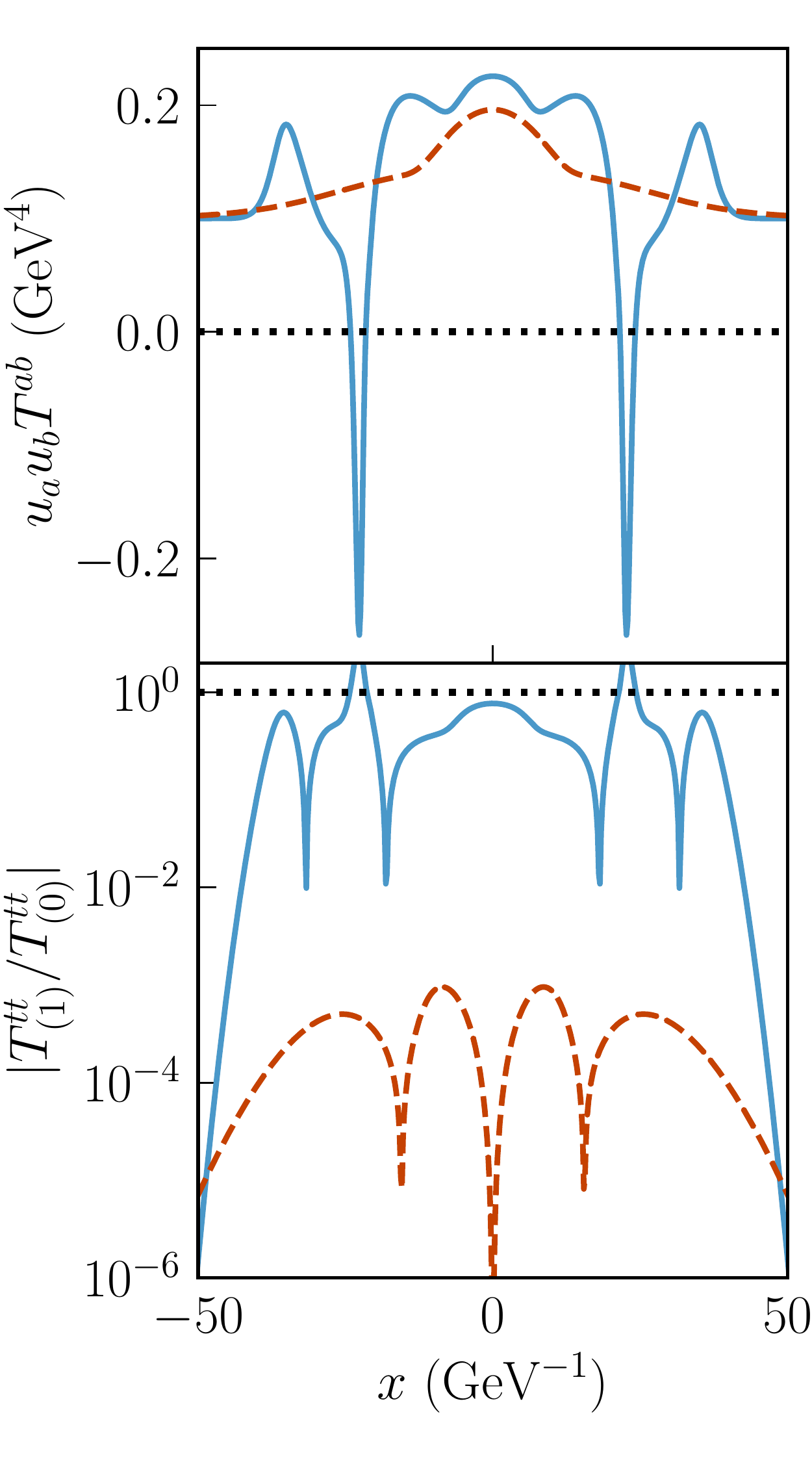}
	\caption{Comparison between the BDNK solution (solid blue line) and MIS solution (dashed red line) for the $\eta/s = 20 \cdot (4 \pi)^{-1}$ case from the rightmost panel of Fig. \ref{fig:FOCS_MIS_comp_smooth}.  The top panel shows that the BDNK solution violates the weak energy condition, $u_{a} u_{b} T^{ab} < 0$, at $x = \pm 23 ~\textnormal{GeV}^{-1}$, while the MIS solution has $u_{a} u_{b} T^{ab} \geq 0$ at all times during the simulation.  Bottom panel: comparison of the BDNK and MIS solutions for the quantity $|T^{tt}_{(1)}/T^{tt}_{(0)}|$. The BDNK solution (solid blue line) has $|T^{tt}_{(1)}/T^{tt}_{(0)}| \geq 1$ at the same place where the weak energy condition is violated.  The MIS solution (dashed red line) stays below $1$ throughout the simulation.} \label{fig:FOCS_MIS_diagnostics}
\end{figure}

The qualitative change in behavior of the BDNK evolution evident in the rightmost panel of Fig.~\ref{fig:FOCS_MIS_comp_smooth} leads one to question if the high viscosity has pushed the system outside of the regime of convergence of the gradient expansion (or at least outside of where only first-order corrections are adequate).  The diagnostics (see Sec. \ref{sec:diagnostics}) shown in Fig. \ref{fig:FOCS_MIS_diagnostics} for this case seem to confirm this suspicion, as the BDNK solution (blue lines) violates the weak energy condition (top panel) and has $|T^{tt}_{(1)}| > |T^{tt}_{(0)}|$ (bottom panel) at certain locations in the flow. 

Interestingly, the MIS solution for the same initial data and viscosity shows no indication (Fig.~\ref{fig:FOCS_MIS_diagnostics}, dashed red line), via the same diagnostics, that one may be in a regime outside the validity of first-order dissipative hydrodynamics.  This occurs because the truncated MIS evolution equation for $\pi^{xx}_{2}$ (\ref{eq:pi_evol_eqn_in_basis}) only includes the Landau frame first-order correction, which has gradients of $v$ but not $\epsilon$, the latter being much more relevant for this particular evolution.  In general, these terms would appear at second order,
and would likely dominate the evolution and give significantly different results from the case shown in Fig. \ref{fig:FOCS_MIS_diagnostics}. This suggests the diagnostics we have considered here are not effective to judge whether one can trust the results of the truncated MIS evolution, and instead one should monitor the magnitude of second-order terms that were dropped\footnote{In the literature (e.g. \cite{Baier_2008,Grozdanov_2016}) it is common to use the zeroth-order equations of motion to simplify the terms at second order and above.  For example, in \cite{Baier_2008} the second-order terms are expressed entirely in terms of $v$, eliminating gradients of $\epsilon$ (though they use $T \propto \epsilon^{1/4}$ as a variable rather than $\epsilon$).  In these cases, one would need to monitor that the zeroth-order EOM are being satisfied to $O(\nabla)$ in order to justify using them to replace $\epsilon$ gradient terms.}.

\begin{figure*}
	\centering
	\includegraphics[width=\textwidth]{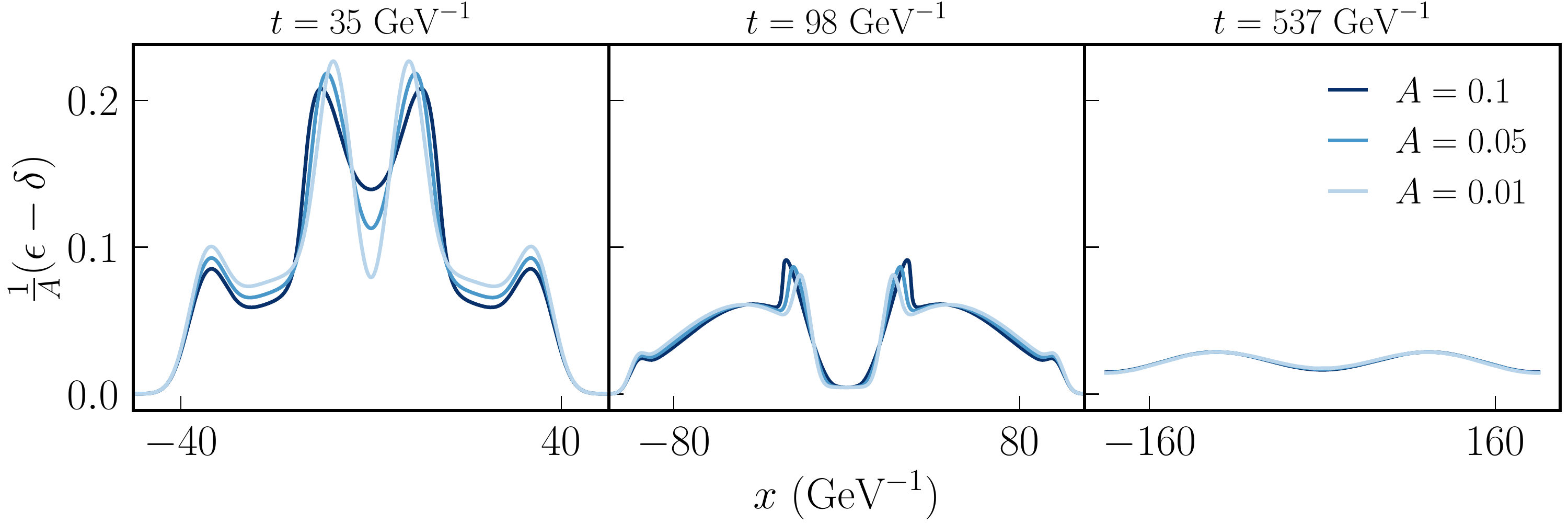}
	\caption{Behavior of smooth BDNK (frame {\tt B}) solutions passing through the `far from equilibrium' phase, as a function of Gaussian amplitude $A$ (cf. (\ref{eq:gaussian_ID})) for $\eta/s = 20 \cdot (4 \pi)^{-1}$.  To aid comparison, what is plotted is the energy density minus the initial background value, $(\epsilon-\delta)$, then scaled by $1/A$; this is done so that all curves overlap at $t = 0$. In all cases, the solution forms a structure with four peaks; these peaks decay at a rate proportional to their amplitude, and the solution eventually settles to one with only two propagating maxima (within the periodic domain).  The late-time solutions for these sets of initial data are very similar between BDNK and MIS.} \label{fig:eps_vs_time}
\end{figure*}

\begin{figure}
	\centering
	\includegraphics[width=\columnwidth]{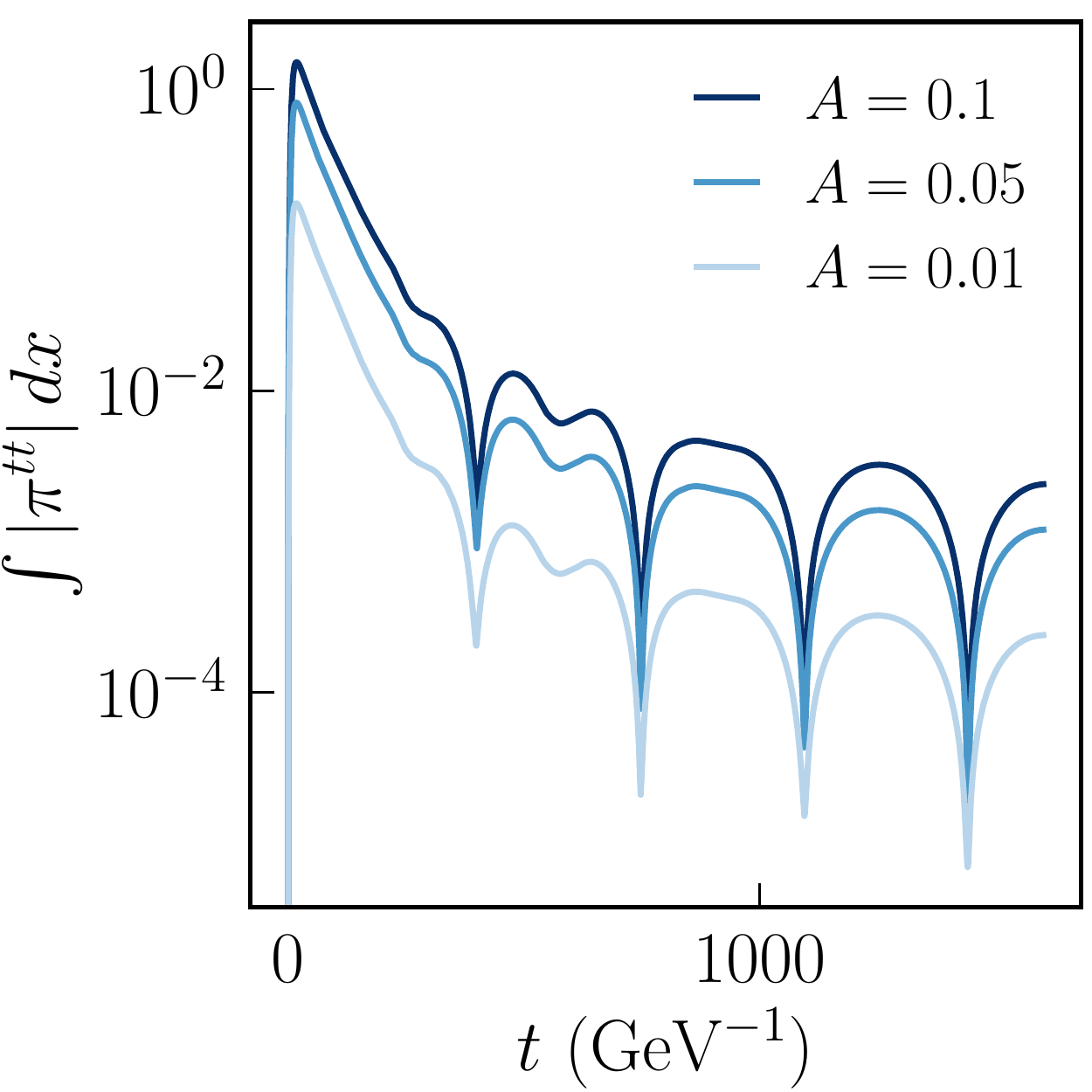}
	\caption{Decay rate of the spatial integral of $|\pi^{tt}|$ over the simulation domain as a function of time---a proxy for the total effect of dissipation on the solution---for the three BDNK (frame {\tt B}) cases at $\eta/s = 20 \cdot (4 \pi)^{-1}$ shown in Fig.~\ref{fig:eps_vs_time}. At early times when in the `far from equilibrium' phase, where $\epsilon$ develops four peaks, the dissipative correction decays exponentially. The end of the exponential phase coincides with these peaks being essentially completely smoothed out, and then there is a transition to a slower power-law decay (the oscillatory features are introduced by the periodic boundary conditions). At these later times,
BDNK (in both frames {\tt A} and {\tt B}) and MIS solutions are in good agreement, despite the qualitative disagreement at earlier times (as can be seen in the right panel of Fig. \ref{fig:FOCS_MIS_comp_smooth}, for example).} \label{fig:pi_decay}
\end{figure}

\begin{figure*}
	\centering
	\includegraphics[width=\textwidth]{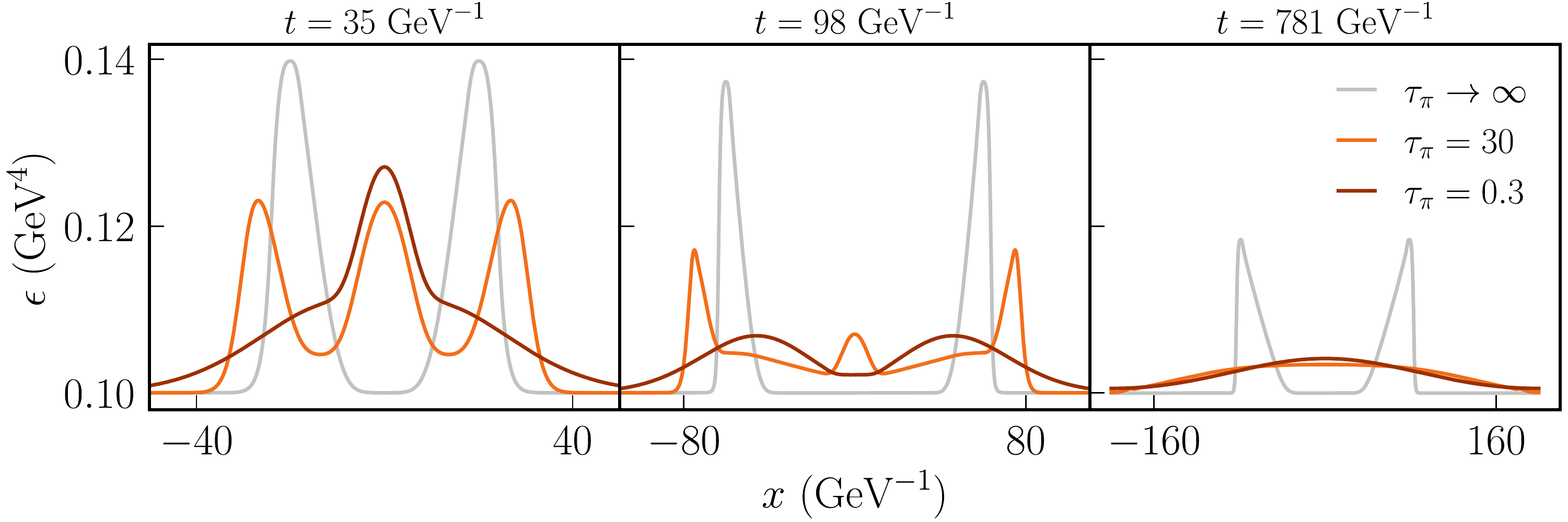}
	\caption{Behavior of smooth MIS solutions for $\eta/s = 20 \cdot (4 \pi)^{-1}$ as a function of the relaxation time $\tau_\pi$, along with the perfect fluid solution $\eta = 0$ (equivalent in this case to the limit $\tau_{\pi} \to \infty$).  The two cases with finite $\tau_{\pi}$ eventually approach a common solution, which agrees with that from the BDNK equations.} \label{fig:eps_vs_tau}
\end{figure*}

As discussed above and illustrated in Figs.~\ref{fig:FOCS_MIS_comp_smooth}-\ref{fig:FOCS_MIS_diagnostics}, 
with large dissipative terms the BDNK vs MIS evolutions become starkly different soon after evolution begins,
and as judged by the BDNK diagnostics are well outside the regime of near-equilibrium
hydrodynamics. Remarkably though, after their initial growth, the large gradients in 
BDNK decay quite rapidly, returning to solutions that are very similar to those obtained 
with MIS, and show no distinctive features left over from this far from equilibrium 
phase---see Fig. \ref{fig:eps_vs_time} for later time
snapshots, and also a comparison between evolutions beginning with different amplitude initial data.
This is reminiscent of so-called universal attractor behavior observed in solutions of various
beyond-ideal theories applied to Bjorken flow ~\cite{Heller:2015dha,Romatschke:2017vte}. There,
essentially arbitrary initial data (within the class relevant to the highly
symmetric Bjorken flow) quickly approaches a hydrodynamic attractor solution
via the decay of non-hydrodynamic modes present in the dissipative theories. 
Though we have not performed any mode analysis in our simulations, this qualitatively
seems to describe what happens here as well; for example, in Fig.~\ref{fig:pi_decay} we
plot norms of $\pi^{tt}$ for the runs depicted in Fig.~\ref{fig:eps_vs_time}, showing
an initial fast exponential decay, followed by a slower power-law decay\footnote{Incidentally,
Fig. \ref{fig:pi_decay} also makes it clear that despite the initial data having $\pi^{ab}(t=0)=0$,
and hence by definition will have the same evolution as the ideal fluid case
precisely at $t=0$, this still constitutes a far-from-ideal initial condition; i.e., we are simply starting
at a zero-crossing of $\pi^{ab}$, which also occurs periodically at later
times due to our periodic domain.}. Presumably
the exponential phase is the decay of the non-hydrodynamic modes, which for BDNK
could be explained (mathematically) as coming from the second-order nature of the PDEs.
Similar behavior should also be present in the MIS evolution, where
the non-hydrodynamic modes can be associated with the treatment of $\pi^{ab}$ as an independent tensor.
This indeed seems to the case, though to make it more evident one needs to
increase the relaxation time parameter $\tau_{\pi}$---see Fig.~\ref{fig:eps_vs_tau}.

\subsection{Discontinuous initial data} \label{sec:subsonic_discont_ID}
A standard test for fluid codes is the so-called shock tube problem: an initially static configuration, but with different constant
energy densities (and pressures) to the left and right of a fictitious membrane separating these states
(at $x=0$ here), that is ``removed'' at $t=0$. 
As discussed earlier, such initial data is mathematically justifiable for the Euler equations, and by extension then the MIS
equations considering $\pi^{xx}_2$ to truly be an independent degree of freedom, but it is unclear
whether similar justification could be made for the BDNK equations. Nevertheless, we compare such evolutions
for the three different theories in this section. Specifically, for our step function discontinuity in $\epsilon$ we choose:
\begin{equation} \label{eq:shock_initial_data}
\epsilon =
\begin{cases}
0.4 ~\textnormal{GeV}^{4} & x < 0  \\
0.1 ~\textnormal{GeV}^{4} & x \geq 0.
\end{cases}
\end{equation}

\begin{figure}
	\centering
	\includegraphics[width=\columnwidth]{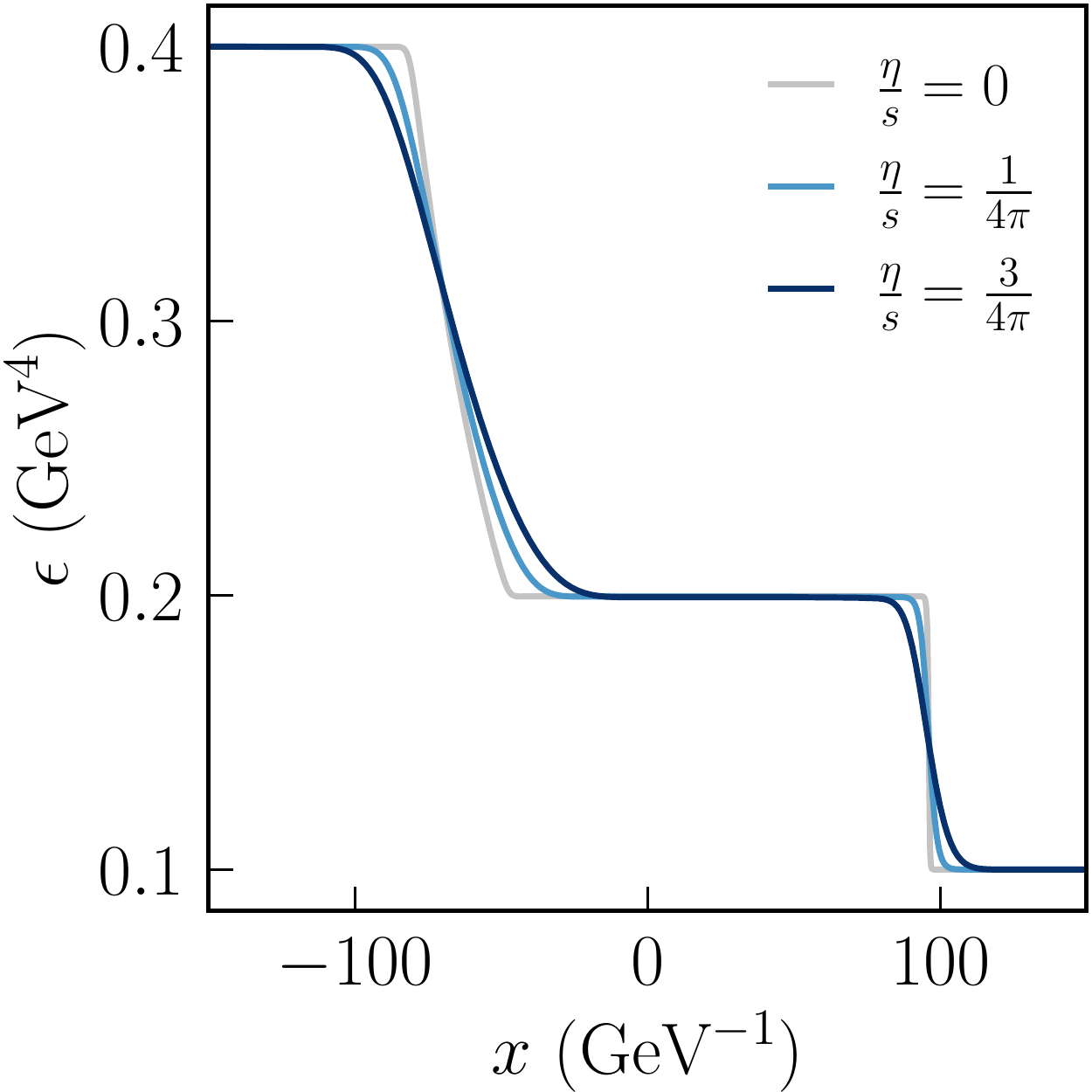}
	\caption{Qualitative effect of viscosity on the evolution of discontinuous initial data for $\eta/s = \{0, 1, 3\} \cdot (4 \pi)^{-1}$ at $t = 35 ~\textnormal{GeV}^{-1}$ (BDNK equations, frame {\tt A}).  Once again, viscosity smooths out the entire profile, including both the rarefaction fan (here at $x \approx -60 ~\textnormal{GeV}^{-1}$) and the forward-propagating shock front (here at $x \approx 100 ~\textnormal{GeV}^{-1}$).} \label{fig:shock_ID_qualitative}
\end{figure}

The qualitative behavior of these solutions is shown in Fig. \ref{fig:shock_ID_qualitative}, again for the relativistic Euler equations ($\eta/s = 0$) and the BDNK equations ($\eta/s = \{1, 3\} \cdot (4 \pi)^{-1}$) using frame {\tt A}, where once again the MIS solutions at these viscosities are nearly identical.  In all cases, three regions form: a backward-propagating rarefaction region, a forward-propagating shock front, and a plateau connecting the two regions.  Dissipation in the BDNK solution smooths out the rarefaction region and the shock front, while in the perfect fluid solution the latter remains discontinuous.  Despite the smoothing, all features propagate at essentially the same speeds.

\begin{figure*}
	\centering
	\includegraphics[width=0.85\textwidth]{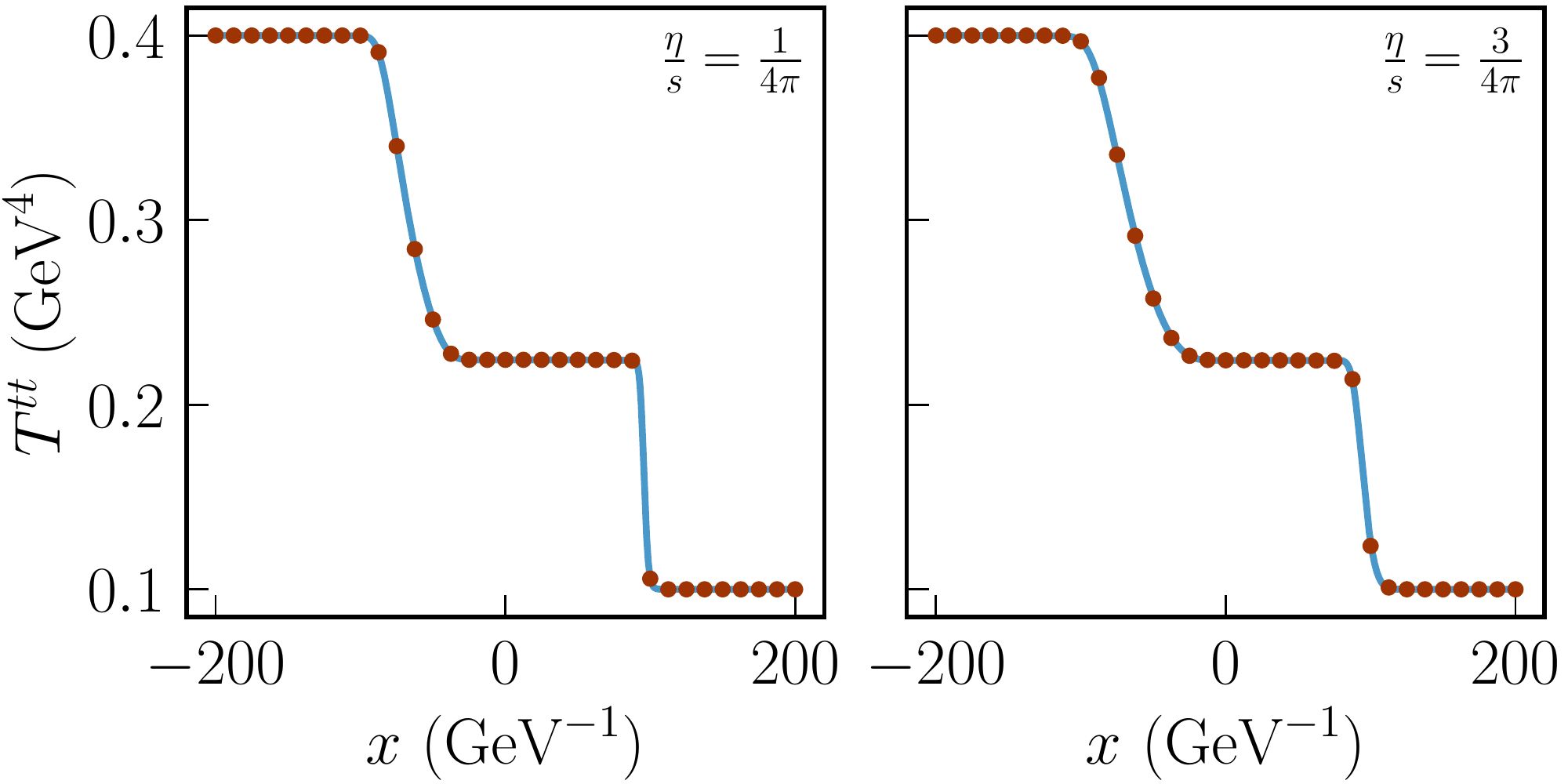}
	\caption{Comparison between the BDNK (frame {\tt A}; blue lines) and MIS solutions (red dots) evolved from discontinuous initial data, here at $t = 35 ~\textnormal{GeV}^{-1}$, for $\eta/s = \{1, 3\} \cdot (4 \pi)^{-1}$. As in Fig. \ref{fig:FOCS_MIS_comp_smooth}, to avoid clutter the MIS points are a sparse sampling of the actual resolution of the simulation.
} \label{fig:FOCS_MIS_comp_shock}
\end{figure*}

In Fig. \ref{fig:FOCS_MIS_comp_shock} we show a comparison between the BDNK and MIS solutions for this initial data, similar to Fig. \ref{fig:FOCS_MIS_comp_smooth}.  As in the smooth data comparison, we find that the BDNK and MIS solutions are effectively identical for $\eta/s = \{1, 3\} \cdot (4 \pi)^{-1}$.  

It is important to note that for discontinuous initial data, our BDNK evolution becomes ``increasingly numerically unstable'' with resolution. By this we mean, as we increase resolution, more ad-hoc numerical ``tricks'' are needed to evolve without a crash at $t=0$; these are, as described in Sec. \ref{sec:RNS_discretization}, treating the perfect fluid flux with a Roe scheme, and adding increasing amounts of Kreiss-Oliger dissipation. With the initial data in (\ref{eq:shock_initial_data}), going
beyond $N=2048+1$ our current algorithm fails. At lower resolutions, we also do not see convergence about $(x=0,t=0)$,
though soon afterward the solutions begin to converge (see Appendix~\ref{sec:convergence_tests}). In that sense then
the solutions shown in Fig.~\ref{fig:FOCS_MIS_comp_shock} can be considered
valid approximate solutions to the BDNK and MIS equations, though (in particular for BDNK) we cannot claim they have evolved from a discontinuity at $t=0$. On the other hand, given the close similarity between the BDNK and MIS solutions, and that these seem to approach the perfect fluid case
as viscosity decreases, suggests a smooth (convergent) approximation to step function initial data would approach this solution in the 
limit for BDNK, even if the exact limiting case is not well defined (and of course, regardless, as the limit is taken beyond some point one
would expect to violate the assumptions of the gradient expansion).

\subsection{Supersonic (shock) initial data} \label{sec:shock_ID}

\subsubsection{Shockwaves in the relativistic Euler equations}

As mentioned earlier, a well-known property of the inviscid equations is that flows which are initially smooth and subsonic can evolve to a state with discontinuities.  While the formation of these discontinuities is nontrivial and not yet fully understood \cite{Pan_2005} \cite{Gremaud_2014}, it is simpler to see why they persist once formed (beyond the intuition that without viscosity there is
no mechanism to smooth them out). This comes from considering the
characteristics of the PDEs, which for the relativistic Euler equations with conformal fluid equation of state, evaluated in the rest frame $v=0$ of the fluid, are
\begin{equation}
\mathfrak{c}_{\pm}^{\textnormal{RE}} = \pm c_s = \pm \frac{1}{\sqrt{3}}.
\end{equation}
This tells us that a supersonic flow (defined by $|v| > c_s$) moves faster than the equations can propagate information, namely
at $\mathfrak{c}_{\pm}^{\textnormal{RE}}$ relative to $v$.  A shockwave is a discontinuity that propagates supersonically, hence there is no way that the 
structure of the subsonic region ahead of a shockwave can inform the structure of the supersonic region behind the shockwave, and the discontinuity
must persist.

The Euler equations thus describe a physical shockwave as a step function discontinuity bridging the upstream and downstream states.
By asserting that this jump in the fluid state satisfies the weak formulation of the 
conservation law (\ref{eq:general_conservation_law}), one arrives at the Rankine-Hugoniot conditions, one of which 
gives the propagation speed $u_s$ of the shock front 
\begin{equation} \label{eq:Rankine_Hugoniot}
u_{s} = \frac{\textbf{f}_{[1]}(x_{L}) - \textbf{f}_{[1]}(x_{R})}{\textbf{q}_{[1]}(x_{L}) - \textbf{q}_{[1]}(x_{R})}.
\end{equation}
Here, the shock is propagating in the $x$ ($i=1$) direction, and 
the components of the flux and state vectors $\textbf{f}_{[1]}$ and $\textbf{q}_{[1]}$, respectively, 
are evaluated just to the left ($x_{L}$) and the right ($x_{R}$) of the shock front. 

In the following subsection we will study propagating shockwaves separating two asymptotic perfect fluid states, $\epsilon_{L}, v_{L}$ at $x \to -\infty$ and $\epsilon_{R}, v_{R}$ at $x \to +\infty$.  In the rest frame of these shocks, the steady-state solution is time-independent, and the relativistic Euler, BDNK, and MIS PDEs reduce to coupled ODEs.  Without time dependence, the Euler equations (\ref{eq:PF_t_eqn})-(\ref{eq:PF_x_eqn}) become $S' = 0, (S v + P)' = 0$, which have nontrivial solutions given by
\begin{equation} \label{eq:PF_jump_conditions}
\begin{aligned}
\epsilon(x), v(x) &= 
\begin{cases}
\epsilon_{L}, v_{L} & x \leq 0 \\
\epsilon_{R}, v_{R} & x > 0
\end{cases} \\
\epsilon_{R} &= \epsilon_{L} \frac{9 v_L^2 - 1}{3 (1 - v_{L}^2)} \\
v_{R} &= \frac{1}{3 v_{L}}.
\end{aligned}
\end{equation}
These are the Rankine-Hugoniot conditions boosted to the reference frame where $u_s=0$.
Hence, considering a flow to the right ($v>0$), after specifying $\epsilon_{L}, v_{L}$, the 
full solution is determined for all $x$, with a step function jump connecting the two asymptotic states at $x \to \pm \infty$.  
Note that restricting to right-moving shockwaves $v_{L} > 0$, non-trivial ($\epsilon_{L} \ne \epsilon_{R}, v_{L} \ne v_{R}$) solutions do exist for $0<v_{L}<1/\sqrt{3}$; however, for $1/3 < v_{L} < 1/\sqrt{3}$ they violate the second law of thermodynamics (the right state has less entropy density than the left), and for $v_{L} < 1/3$ the right state is superluminal and has negative energy density. 
Thus right-moving physical shockwaves only exist for $v_{L}>1/\sqrt{3}$.

Since we are considering a shockwave joining two asymptotic {\em equilibrium} states, the solutions for viscous fluids, considered in the next section, should be well approximated by (\ref{eq:PF_jump_conditions}) outside a finite region around the shockwave itself (or said another way, the viscous solutions will replace what is a step
function solution of the Euler equations with a smooth transition between the same asymptotic end-states).

\subsubsection{Shockwaves in viscous fluids}\label{sec:shock_ID_vis}

One is forced to accept discontinuous shockwave solutions to the relativistic Euler equations because all shockwaves propagate faster than the characteristic speeds of the equations.  This behavior is not shared by the BDNK and MIS equations, as they have a larger  number of characteristic speeds, some of which are greater than the fluid sound speed.  This allows for the possibility that these theories can possess continuous shock solutions. Such solutions have been investigated for certain MIS-type theories, where they were found to exist only so long as the upstream flow velocity is less than the maximum characteristic speed of the system~\cite{Olson_1990_shocks,Geroch_1991}. 

Guided by these results, we apply similar reasoning to the two viscous theories considered here. The first step is to compute the characteristic speeds of the PDEs we evolve. For the BDNK equations the result is (again for simplicity expressed in the rest frame $v=0$ of the fluid):
\begin{equation}\label{eq:BDNK_threshold_v}
\mathfrak{c}_{i}^{\textnormal{BDNK}} = \pm \sqrt{\frac{\chi_0 (2 \eta_0 + \lambda_0) \pm 2 \sqrt{\eta_0 \chi_0 (\chi_0 (\eta_0 + \lambda_0) + \lambda_0^2)}}{3 \lambda_0 \chi_0}}.
\end{equation}
Notice that this expression depends on all the first-order transport coefficient parameters, which, crucially, depend on the hydrodynamic frame.
For frame {\tt A} (\ref{frames}), (\ref{eq:BDNK_threshold_v}) evaluates to 
\begin{equation}\label{eq:BDNK_cs1}
\mathfrak{c}_{i}^{\textnormal{BDNK,A}} = \pm \sqrt{\frac{31 \pm 2 \sqrt{134}}{75}} \sim \pm 0.32, \pm 0.85.
\end{equation}
That the maximum speed is less than the speed of light implies (and as we show empirically is true, and also recently
proven in~\cite{freistuhler2021nonexistence} in an independent work),
that arbitrarily strong,
smooth shock solutions do not exist within this frame. This inspired us to consider frame {\tt B} (\ref{frames}), where we chose
the frame parameters specifically so that the maximum speed is equal to the speed of light (this is not the unique choice, but is a 
particularly simple example):
\begin{equation}\label{eq:BDNK_cs2}
\mathfrak{c}_{i}^{\textnormal{BDNK,B}} = \pm 1, \pm \frac{1}{5}.
\end{equation}

Performing the same calculation for the truncated MIS equations, one finds (again with $v=0$) 
\begin{equation}\label{eq:MIS_cs}
\mathfrak{c}_{i}^{\textnormal{MIS}} = 0, \pm \sqrt{\frac{1}{3} + \frac{4 \eta_0 \epsilon^{3/4}}{(4 \epsilon + 3 \pi^{xx}_{2}) \tau_{\pi}}}.
\end{equation}
One can think of the zero-speed mode as being associated with the transport equation for $\pi^{xx}_2$ (\ref{eq:pi_evol_eqn_in_basis}),
and the other two giving the characteristic speeds of the fluid variables.
Notice that, in contrast to the BDNK characteristics above, the nonzero speeds 
do depend on the state of the system, which underlies the claims~\cite{Olson_1990_shocks,Geroch_1991} that 
MIS-type theories do not allow strong shock solutions in all situations; i.e., one can always find some state where the maximum characteristic speeds are less than 1 (in (\ref{eq:MIS_cs}) for sufficiently large $\epsilon$, for example).
In contrast, it is easy to choose values for $\tau_{\pi}$ such that the characteristic speeds are superluminal, and in fact,
that is the case for the shockwave examples discussed below when using MIS, as well as most other
cases presented here using $\tau_{\pi}=0.3 ~\textnormal{GeV}^{-1}$. However,
with these, and all other examples we have looked at, the solutions do not seem to exhibit any problematic behavior; i.e.
the equations ``merely'' happen to have characteristic-cones that lie outside the light cone. In particular, near equilibrium,
localized perturbations in the fluid still propagate at the sound speed, and when far from equilibrium,
the dynamics, though much more complicated as illustrated in Fig.~\ref{fig:eps_vs_tau}, still do
not seem to show superluminal propagation of prominent features, nor flow velocities that become
superluminal. Thus it is unclear under what circumstances
a superluminal characteristic structure leads to violation of causality in the problematic
sense of the phrase. 
For a detailed analysis of this issue for the wider class of MIS theories, see \cite{Bemfica_2020_MIS,Plumberg:2021bme}.

To numerically explore shockwave solutions, for initial data we choose the following smooth transition between 
two chosen asymptotic states ($\epsilon_{L}, v_{L}>1/\sqrt{3}$) and ($\epsilon_{R}, v_{R}<1/\sqrt{3}$):
\begin{equation} \label{eq:smooth_shock_ID}
\begin{aligned}
\epsilon(x, t=0) &= \frac{(\epsilon_{R} - \epsilon_{L})}{2} \Big[ \textnormal{erf} \Big( \frac{x}{w} \Big) + 1 \Big] + \epsilon_{L} \\
v(x, t=0) &= \frac{(v_L-v_R)}{2} \Big[ 1 - \textnormal{erf} \Big( \frac{x}{w} \Big) \Big] + v_R,
\end{aligned}
\end{equation}
where $\textnormal{erf}(x/w)$ is the Gaussian error function\footnote{One can show that this set of initial data approaches a step function jump in state at $x=0$ in the limit $w \to 0^{+}$ using the identity \cite{Bracewell_1965} $\lim\limits_{w \to 0^{+}} \textnormal{erf}\left(\frac{x}{w}\right) = 2 \Theta(x) - 1$, where the step function $\Theta(x > 0) = 1$ and $\Theta(x \leq 0) = 0$.}.
For the examples shown here we set $w=10$, freely choose $\epsilon_{L}, v_{L} >1/\sqrt{3}$, and then compute $\epsilon_{R}, v_{R}$ using the perfect fluid jump conditions (\ref{eq:PF_jump_conditions}). 

\begin{figure}
	\centering
	\includegraphics[width=\columnwidth]{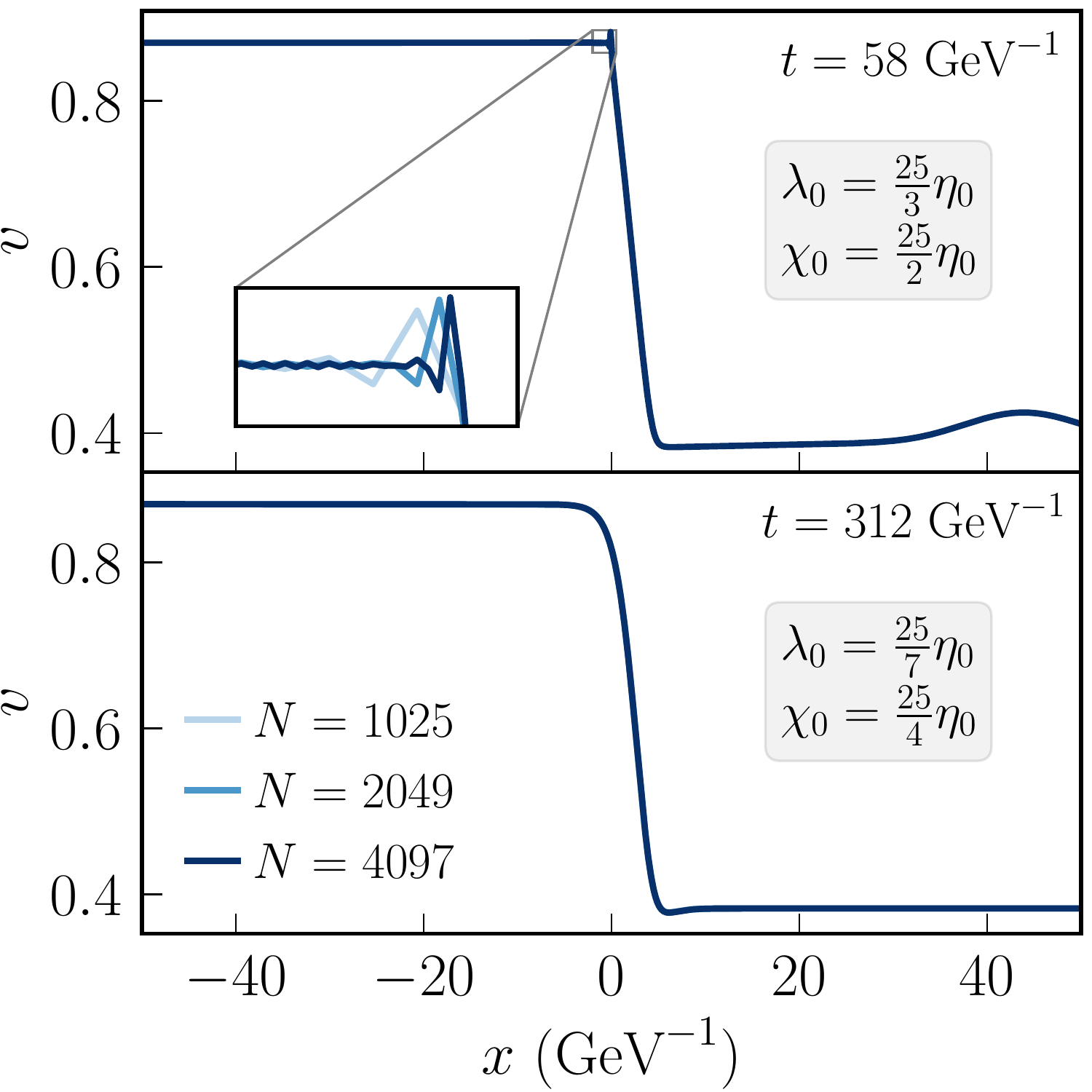}
	\caption{BDNK evolution of supersonic shock initial data (\ref{eq:smooth_shock_ID}) with $\epsilon_{L} = 1, v_{L} = 0.87$, and
$\epsilon_{R},v_{R}$ given by the perfect fluid jump conditions (\ref{eq:PF_jump_conditions}).  The top panel was obtained using frame {\tt A} (\ref{frames}), which has a maximum rest-frame characteristic speed (\ref{eq:BDNK_cs1}) less than $v_L$. This results in an instability that causes the code to crash soon after the time depicted. The inset focuses in on where the instability first develops. Overlaid are the results from three different resolution runs; the higher resolutions (darker curves) crash sooner, indicative of a high frequency numerical instability. Also evident on the top (main) panel near the right edge is the transient ``blob'' mentioned in the text, which is an artifact of the initial data not matching the stationary shock profile between the two chosen end states. The bottom panel is the same initial data obtained with frame {\tt B} that has a maximum characteristic speed equal to the speed of light (\ref{eq:BDNK_cs2}). No instability occurs, and a steady state is reached (notice the much later time stamp, in particular long after the transient blob has propagated off the domain).} \label{fig:frame_diff}
\end{figure}

We find that, evolving with the BDNK equations, all members of the family of initial data (\ref{eq:smooth_shock_ID}) we studied converge to smooth, 
steady-state shock profiles as long
as $v_L$ is less than the maximum characteristic speed of the particular frame (which for frame
{\tt B} (\ref{eq:BDNK_cs2}) includes all cases up to the largest velocities our code can generically handle). The typical evolution
for such a case begins with a transient ``blob'' of fluid forming around the transition, that then propagates off to the right, after which the fluid relaxes to the steady-state profile\footnote{If one does not choose the right state conditions to match the perfect fluid ones (\ref{eq:PF_jump_conditions}), the transient feature that propagates to the right is correspondingly larger, and the solution can settle to a steady state where the shock front is moving in the simulation reference frame. Boosting to the rest frame of this shock front then gives a solution that does satisfy (\ref{eq:PF_jump_conditions}) asymptotically.}.
For frame {\tt A} (\ref{eq:BDNK_cs1}),
when $v_L\gtrsim 0.85$, some time after evolution begins a high frequency instability develops near the left side of the shock
transition---see Fig. \ref{fig:frame_diff}, where we also show the same case obtained with frame {\tt B} for comparison.

\begin{figure}
	\centering
	\includegraphics[width=\columnwidth]{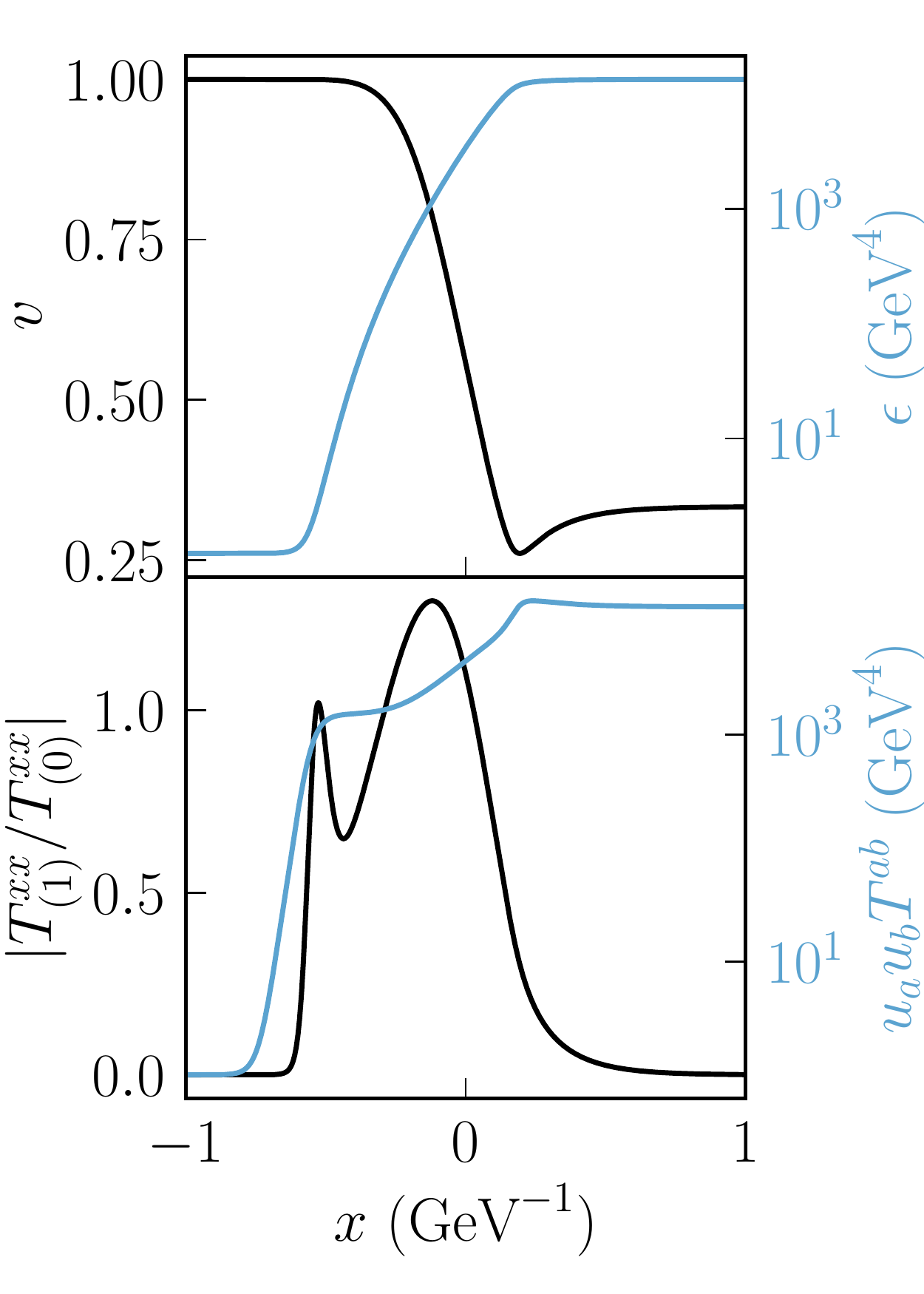}
	\caption{Shockwave solution to the steady-state BDNK equations (\ref{eq:BDNK_ODE_epsprime}-\ref{eq:BDNK_ODE_vprime}) in frame {\tt B} (\ref{frames}), with $v_{L} = 0.9999, \epsilon_L=1$.  Top panel: solution in $v$ (black) alongside that in $\epsilon$ (blue), the latter on a log scale because $\epsilon_{R} \to \infty$ as $v_{L} \to 1$.  Note that these solutions are qualitatively similar to those for smaller $v_{L}$ (see Fig. \ref{fig:frame_diff} for example), except for the differences in scale. Bottom panel: measures of the gradient expansion convergence $|T^{xx}_{(1)}/T^{xx}_{(0)}|$ (black) and the weak energy condition $u_{a} u_{b} T^{ab}$ (blue) for the same solution.  Notice that the former exceeds unity in the central transition region, implying the solution is outside the regime of validity of the gradient expansion here (the less relativistic the shock the smaller the maximum of $|T^{xx}_{(1)}/T^{xx}_{(0)}|$ becomes; for example, it is about an order of magnitude smaller for the case shown in Fig.~\ref{fig:frame_diff}). The weak energy condition is not violated anywhere.
} \label{fig:ODEs}
\end{figure}

Since our code develops other numerical problems when flow speeds are larger than $v\sim 0.9$, it is reasonable
to question whether we can indeed claim that the BDNK equations allow smooth shockwaves for arbitrarily large
upwind speeds (note that the recent proof~\cite{freistuhler2021nonexistence} of failure of existence
of sufficiently strong shock solutions in strictly causal frames, as frame {\tt A}, does not prove that in sharply 
causal frames, as frame {\tt B}, that {\em all} shocks must have smooth profiles). 
To give further evidence for this claim, we note that the steady state solutions we evolve to in the dynamical
code match, to within truncation error, the ``exact'' stationary profiles one can obtain by integrating
the ODEs governing the corresponding time-independent limit of the BDNK equations (listed in Appendix~\ref{app_ss}).
Moreover, these ODEs have singularities (not coincidentally) at {\em exactly} the points where the flow velocity
crosses a characteristic speed of the system. Also, from 
the ODEs one can estimate that the characteristic width of the transition from left-to-right asymptotic states, when a solution exists,
scales as $\sim (1-v_L^2)^{1/4}$; hence, the steepening of the shock profile with strength $v_L$ is largely frame {\em independent},
and a discontinuity only forms in 
the limit $v_L\rightarrow 1$ (though above some value of $v_L$ before $1$, the gradients in the transition region will become large enough that 
one would not trust the first-order theory to give an accurate description of the shock profile there).
To illustrate, in Fig.~\ref{fig:ODEs}, we show a solution to the ODEs for $v_L=0.9999$,
as well as plots of the diagnostics vetting the first-order description.

For the MIS equations we evolved the same set of cases as with the BDNK examples; all MIS evolutions
had maximum characteristic speeds greater than 1
throughout the domain, and reached similar, stable steady-states. 

\section{Conclusion} \label{sec:conclusion}

We have performed a first numerical study of a class of causal, stable, first-order relativistic hydrodynamic theories recently developed
by Bemfica, Disconzi, Noronha~\cite{Bemfica_2018} and Kovtun~\cite{Kovtun_2019}.
The results are encouraging in that for smooth initial data, with small to moderate
viscosity, the results agree with those obtained by a code based on the 
M\"uller-Israel-Stewart formalism commonly used today when dissipative
effects are important in a relativistic setting. The latter requires
appealing to second-order effects to circumvent problems that arise
when traditional (Eckart or Landau-Lifshitz) hydrodynamic frames
are adopted.  Though this approach works, it is reassuring and could prove
more useful in certain situations that one can instead self-consistently
and stably remain within the realm of first-order hydrodynamics if the corresponding 
dissipative effects are adequate to model the problem at hand.

Regarding larger viscosities, an appealing feature of the first-order
theories is they offer simple diagnostics that can be 
used to judge whether a given flow is adequately described by first-order
only dissipative effects (in contrast to the truncated-MIS evolutions considered
here, which would have required computation of dropped second-order terms to realize
that the solutions were outside of the regime of validity, from the perspective
of a gradient expansion). However, for the cases we explored,
even when starting with initial data with large gradients (far from equilibrium), 
the evolution quickly carried the system back to the hydrodynamic (small-gradient) regime,
and---quite remarkably---did so in a manner that seemed to erase all signs of the non-hydrodynamic
behavior in the flow that developed at early times. This is similar to universal attractor
behavior found in Bjorken-like flows ~\cite{Heller:2015dha,Romatschke:2017vte},
though here in cases with less symmetry.

Our results on strong shockwave solutions, consistent with the recent work~\cite{freistuhler2021nonexistence},
also suggests that arbitrarily
strong, smooth shocks are generically allowed in the first-order relativistic theories,
if an appropriate class of hydrodynamic frames are employed.
This is another possible advantage over MIS-type 
theories, which do not share this feature~\cite{Olson_1990_shocks,Geroch_1991}.
On the other hand, this might simply suggest that the question 
of the existence of smooth, strong shock solutions in second-order theories needs to
be reconsidered after restoring full freedom to choose the hydrodynamic frame.

Regarding discontinuous initial data, though (as discussed at length earlier in the
paper) it is not clear such data is well-posed within BDNK theory, nevertheless,
when our scheme can stably evolve past $t=0$, the late time solutions agree
well with corresponding MIS solutions, giving 
smoothed versions of the solutions found in the perfect fluid limit (as one would intuitively
 expect). 

There are numerous avenues for follow up work. Within scenarios
where symmetries can reduce the problem to (1+1)D PDEs, as here,
a couple of such directions are to go beyond conformal fluids, and
to include gravity in a spherically symmetric setting.
Relaxing symmetries, it would be interesting to attempt to tackle 
essentially all applications mentioned in the introduction where relativistic,
first-order dissipative effects need to be modeled.

\begin{acknowledgments}
We thank Elias Most for useful conversations related to this work, and Jorge Noronha for useful comments on an earlier draft of the manuscript. This material is based upon work supported by the National Science Foundation (NSF) Graduate Research Fellowship Program under Grant No. DGE-1656466. Any opinions, findings, and conclusions or recommendations expressed in this material are those of the authors and do not necessarily reflect the views of the National Science Foundation. F.P. acknowledges support from NSF Grant No. PHY-1912171, the Simons Foundation, and the Canadian Institute For Advanced Research (CIFAR).
\end{acknowledgments}

\appendix

\section{BDNK primitive variable recovery} \label{sec:RNS_prim_var_recovery}
For the BDNK equations (\ref{eq:general_conservation_law}), (\ref{eq:RNS_conserv_form}), we recover the primitive variables $\bm{p}^{NS} = (\dot{\epsilon}, \dot{v})^{T}$ from the conservative variables $\bm{q}^{NS} = (\pi^{tt}_{1}, \pi^{tx}_{1})^{T}$ analytically via
\begin{multline} \label{eq:RNS_epsdot}
\dot{\epsilon} = -\frac{2}{K} \Big( \epsilon' \left(v^3 \left(2 \eta _0 \lambda _0+6 \eta _0 \chi _0+\lambda _0 \chi _0\right)-3 \lambda _0 v \chi _0\right) \\
+2 \lambda _0 v' \epsilon  \left(v^2 \left(\chi _0-4 \eta _0\right)-3 \chi _0\right) \Big) \\
-\frac{4 \pi^{tt}_{1} \epsilon^{1/4} \left(3 \lambda _0+v^2 \left(-4 \eta _0+3 \lambda _0+4 \chi _0\right)\right)}{K W} \\
+ \frac{4 \pi^{tx}_{1} v \epsilon^{1/4} \left(3 \left(2 \lambda _0+\chi _0\right)+v^2 \left(\chi _0-4 \eta _0\right)\right)}{K W}
\end{multline}
\begin{multline} \label{eq:RNS_vdot}
\dot{v} = -\frac{3 \epsilon' \lambda _0 \left(v^2-3\right) \chi _0}{4 K W^4 \epsilon } \\
+\frac{3 \lambda _0 v \left(\pi^{tt}_{1} v^2 + \pi^{tt}_{1} - 2 \pi^{tx}_{1} v\right)-3 \chi _0 \left(\pi^{tx}_{1} \left(v^2+3\right) - 4 \pi^{tt}_{1} v\right)}{K W^3 \epsilon ^{3/4}} \\
-\frac{2 v v' \left(2 \eta _0 \left(\lambda _0 v^2+3 \chi _0\right)+\lambda _0 \left(v^2-3\right) \chi _0\right)}{K}
\end{multline}
where we have defined the shorthand 
\begin{equation*}
K \equiv -9 \lambda _0 \chi _0+\lambda _0 v^4 \left(4 \eta _0-\chi _0\right)+6 v^2 \chi _0 \left(2 \eta _0+\lambda _0\right).
\end{equation*}
The equations above are regular as long as $\epsilon > 0, v \in (-1, 1)$; 
for both frames (\ref{frames}) considered here the only physical root of $K$ (with $|v|<1$) is $\eta_0 = 0$.

\section{Detailed description of the BDNK algorithm} \label{sec:RNS_pseudocode}
We advance the solution forward in time using Heun's method (\ref{eq:Heuns_method}): beginning from time $t^{n}$ when the state of the fluid ($T^{ab}_{1}$) is completely known, we first evolve to a predictor level $\bar{t}^{n+1}$ before updating to the advanced time $t^{n+1} = t^{n} + \Delta t$.  Henceforth we will denote quantities at the known level with an upper index $n$, quantities at the predictor level with a bar and index $n+1$, and quantities at the advanced level with index $n+1$ and no bar, e.g. we evolve $\epsilon^{n}\to \bar{\epsilon}^{n+1} \to \epsilon^{n+1}$.  

We do this by solving four equations: (\ref{eq:RNS_epsdot}) for $\epsilon$; (\ref{eq:RNS_vdot}) for $v$; and the two components of the conservation law (\ref{eq:general_conservation_law})-(\ref{eq:PF_conserv_terms}), (\ref{eq:RNS_conserv_form})-(\ref{eq:dissipative_flux}) for $\pi^{tt}_{1}, \pi^{tx}_{1}$.  The actions performed for each time integration step of this algorithm are as follows.
\begin{enumerate}
	\item Given $\epsilon, v, \pi^{tt}_{1}, \pi^{tx}_{1}$ are known at $t^{n}$, compute $(\pi^{xx})^{n}$ from its definition (\ref{eq:first_order_const_in_basis})-(\ref{eq:first_order_const_in_basis_b}).
	\item Compute $\dot{\epsilon}^{n}, \dot{v}^{n}$ using (\ref{eq:RNS_epsdot}), (\ref{eq:RNS_vdot}) respectively.  These quantities may be used immediately to compute $\bar{\epsilon}^{n+1}, \bar{v}^{n+1}$, e.g. $\bar{\epsilon}^{n+1} = \epsilon^{n} + \Delta t \, \dot{\epsilon}^{n}$.
	\item Use the two components of the conservation law (\ref{eq:general_conservation_law})-(\ref{eq:PF_conserv_terms}), (\ref{eq:RNS_conserv_form})-(\ref{eq:dissipative_flux}) to compute $\overline{(\pi^{tt}_{1})}^{\,n+1}, \overline{(\pi^{tx}_{1})}^{\,n+1}$, respectively.
	\item Insert $\bar{\epsilon}^{n+1}, \bar{v}^{n+1}, \overline{(\pi^{tt}_{1})}^{\,n+1}, \overline{(\pi^{tx}_{1})}^{\,n+1}$ in (\ref{eq:RNS_epsdot}), (\ref{eq:RNS_vdot}) to compute $\bar{\dot{\epsilon}}^{n+1}, \bar{\dot{v}}^{n+1}$.
	\item Compute $\epsilon^{n+1}, v^{n+1}$ using the second step of (\ref{eq:Heuns_method}).  For example, $\epsilon$ would be evolved via ${\epsilon^{n+1} = \epsilon^{n} + \frac{\Delta t}{2} (\dot{\epsilon}^{n} + \bar{\dot{\epsilon}}^{n+1})}$.
	\item Solve for $(\pi^{tt}_{1})^{n+1}, (\pi^{tx}_{1})^{n+1}$ using the conservation law (\ref{eq:general_conservation_law})-(\ref{eq:PF_conserv_terms}), (\ref{eq:RNS_conserv_form})-(\ref{eq:dissipative_flux}) with values at the known level $t^{n}$ and the predictor level $\bar{t}^{n+1}$ via the second step of (\ref{eq:Heuns_method}).
\end{enumerate}
In steps 3 and 6 above, we optionally use the Roe flux to compute ${\bf f}^{PF}$ (\ref{eq:RNS_conserv_form}) (otherwise we 
use finite differences), and optionally apply Kreiss-Oliger dissipation 
to ${\bf q}^{NS}$ (\ref{eq:RNS_conserv_form}).

\section{BDNK steady-state ODEs}\label{app_ss}
If one restricts to time-independent (steady-state) solutions, all time derivative terms vanish and the PDEs describing stress-energy conservation reduce to coupled ODEs.  The structure of (\ref{eq:T_ab_conservation_law}) further implies (in planar symmetry in Minkowski spacetime) that the remaining equations are total $x$ derivatives of the form $\partial_{x} T^{xb} = 0$, which may be trivially integrated to yield $T^{tx} = C_{1}, T^{xx} = C_{2}$ for real constants $C_{1}, C_{2}$.  For the perfect fluid, the steady-state equations are $S = C_{1}, S v + P = C_{2}$, with a trivial solution $\epsilon, v = \textnormal{constant}$ and a nontrivial solution describing a shockwave given by the Rankine-Hugoniot conditions (\ref{eq:PF_jump_conditions}).

For BDNK theory (\ref{eq:first_order_const_in_basis})-(\ref{eq:BDNK_eqns}), the equations after the trivial integral are coupled nonlinear ODEs; these ODEs may be rearranged to yield
\begin{multline} \label{eq:BDNK_ODE_epsprime}
\epsilon' = \frac{-4 \epsilon^{1/4} \sqrt{1-v^2}}{9 \lambda_0 \chi_0 (v-\mathfrak{c}_{1})(v-\mathfrak{c}_{2})(v-\mathfrak{c}_{3})(v-\mathfrak{c}_{4})} \\
\times \Big( C_1 \left(4 \eta _0-\chi _0\right) + v \big(C_2 \left(-4 \eta _0+3 \lambda _0+4 \chi _0\right) \\
+ 3 \lambda _0 v^2 \left(C_2+\epsilon \right) - 3 C_1 v \left(2 \lambda _0+\chi _0\right)-\epsilon  \left(4 \eta _0+\lambda _0\right)\big) \Big)
\end{multline}
and
\begin{multline} \label{eq:BDNK_ODE_vprime}
v' = \frac{-\left(1-v^2\right)^{3/2} \epsilon^{-3/4}}{9 \lambda_0 \chi_0 (v-\mathfrak{c}_{1})(v-\mathfrak{c}_{2})(v-\mathfrak{c}_{3})(v-\mathfrak{c}_{4})} \\
\times \Big(9 C_1 v^3 \chi _0-3 v^2 \left(C_2 \left(\lambda _0+4 \chi _0\right)+\lambda _0 \epsilon \right) \\
+3 C_1 v \left(2 \lambda _0+\chi _0\right)+\lambda _0 \left(\epsilon -3 C_2\right) \Big),
\end{multline}
where $\mathfrak{c}_{i}$ with $i \in \{1,2,3,4\}$ are the four characteristic speeds from (\ref{eq:BDNK_threshold_v}).  Notice that the equations share the same denominator, which becomes singular when the flow velocity crosses any of $\mathfrak{c}_{i}$.
Thus, as discussed in detail in Sec.~\ref{sec:shock_ID}, a judicious choice of frame is required to be able to represent all steady state solutions of interest, which in a dynamical setting seems to translate to the well-posedness of the corresponding initial value
problem near such states.

\section{Convergence tests} \label{sec:convergence_tests}

\begin{figure}
	\centering
	\includegraphics[width=\columnwidth]{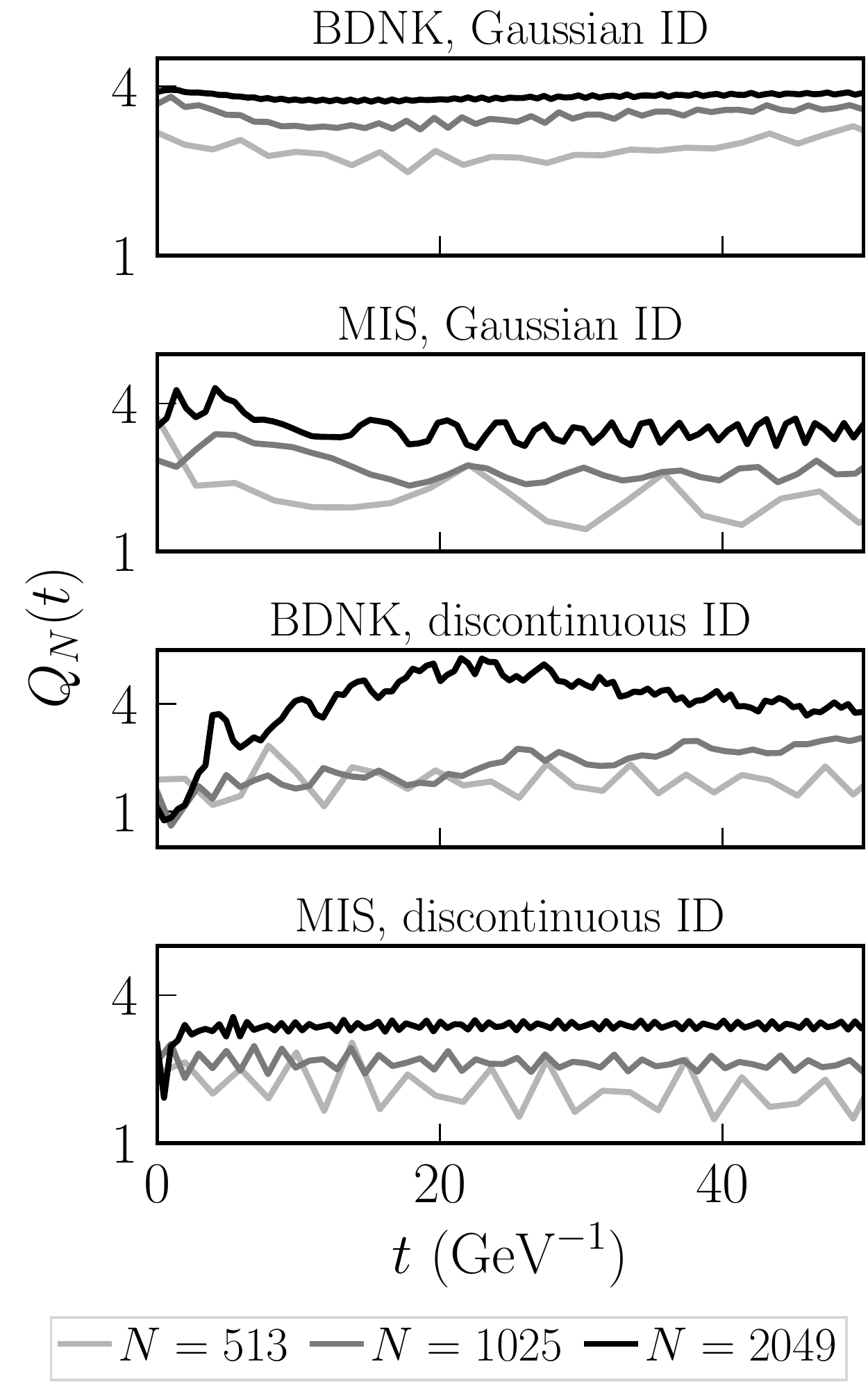}
	\caption{Convergence plots $Q_{N}(t)$ for the BDNK (frame {\tt A}) and MIS solutions for the case $\eta/s = (4 \pi)^{-1}$ for an independent (leapfrog) discretization of the $t$ component of (\ref{eq:T_ab_conservation_law}).  In order of increasing darkness, the lines correspond to $N = 513, 1025, 2049$.  The top two panels correspond to cases with Gaussian initial data (\ref{eq:gaussian_ID}), and show the expected trend to convergence as resolution is increased (in particular, for the BDNK equations this should be second 
order, $Q_{N}(t) \sim 4$, and for the MIS equations somewhere between first and second order, $Q_{N}(t) \sim 2-4$, depending
on how significant the first order advection term is in the solution). The bottom two panels correspond to the discontinuous shock tube initial data (\ref{eq:shock_initial_data}), and do not show convergence near $t = 0$ as measured by $Q_N(t)$; as discussed in the text, this is expected, and once viscosity smooths out the discontinuity we do see return to convergence. } \label{fig:conv_plot}
\end{figure}

For all of the runs performed here, we check for both the correctness of our results and convergence by 
monitoring the rate $Q_N(t)$ at which an independent residual
of the evolution equations (typically the $t$ component of (\ref{eq:T_ab_conservation_law})) converges to zero; specifically 
\begin{equation} \label{eq:Q_factor}
Q_{N}(t) = \frac{||L^{2 h} u^{2 h}||}{||L^{h} u^{h}||},
\end{equation}
where $L^{h} u^{h}$ denotes the discretization of the residual operator $L$ acting on a PDE solution $u$ evolved on a mesh with grid spacing $h = (x_{max}-x_{min})/(N-1)$, and $||\cdot||$ denotes any vector norm; here we use the 1-norm.  The convergence factor (\ref{eq:Q_factor}) divides the discrete residual of a solution computed with grid spacing $2h$ by that computed at spacing $h$, and for smooth solutions can be shown to asymptote to $Q \to 2^{n}$ in the continuum limit $h \to 0$ for a convergent numerical scheme with truncation error $O(h^{n})$.

For all three of the systems of PDEs considered here---the relativistic Euler, BDNK, and MIS equations---all of the discrete elements in the algorithm are second-order accurate, with two exceptions: 
first is that the perfect fluid part of the flux $\bm{f}^{PF}$, as a result of the slope limiter, converges at second order only in regions where the solution is smooth, elsewhere it is first order; the other is in the MIS $\pi^{xx}_{2}$ evolution equation (\ref{eq:pi_evol_eqn_in_basis}), which uses a first-order upwind discretization for the advection operator (\ref{eq:upwind_discretization}).

As a result---see Fig. \ref{fig:conv_plot} for examples---we find that $Q_{N}(t)$ tends to $4$ with increasing resolution for the relativistic Euler and BDNK equations at times when the solution is smooth.
For MIS solutions, though strictly speaking in the limit $h\rightarrow 0$ the first-order term should dominate,
at the resolutions considered here ($N$ from $128+1$ to $2048+1$) we see somewhere between first ($Q_{N}(t)\sim 2$) and second
order ($Q_{N}(t)\sim 4$) convergence. For solutions about discontinuities (\ref{eq:Q_factor}) is not justified, and we do not expect 
(and do not see) convergence by this measure. For the Euler equations, we have checked that we do converge to solutions about
shock fronts that are consistent with the Rankine-Hugoniot conditions (\ref{eq:Rankine_Hugoniot}). As discussed in the main
text, for the BDNK and MIS equations we have not found situations where discontinuities dynamically form, and so the only examples
we looked at are the shock tube tests where we put them in by hand at $t=0$, whether that is justifiable in a weak-sense,
as they are for the Euler equations, or not. Though at least the way the code ``resolves'' these discontinuities,
once some dissipation with evolution has occurred, is consistent with energy-momentum conservation, in particular
in that the resulting smooth shock fronts have the same propagation speeds and asymptotics as in the perfect fluid limit.

\bibliography{references}

\end{document}